\documentclass[aps,twocolumn,showpacs,preprintnumbers,amsmath,amssymb,floatfix]{revtex4}

\usepackage{graphicx}
\usepackage{dcolumn}
\usepackage{bm}
\usepackage[german,american]{babel}

\begin{document}


\title{Does the potential energy landscape of a supercooled liquid resemble a collection of traps?}

\author{A. Heuer$^*$}
 \author{B. Doliwa$^+$}
\author{A. Saksaengwijit$^*$}%

\affiliation{
(*)Westf\"{a}lische Wilhelms-Universit\"{a}t M\"{u}nster, Institut f\"{u}r Physikalische Chemie\\
and International Graduate School of Chemistry\\
Corrensstr. 30, 48149 M\"{u}nster, Germany\\
(+)TU Darmstadt, TEMF, 64289 Darmstadt, Germany\\
}

\date{\today}

\begin{abstract}

It is analyzed whether the potential energy landscape of a
glass-forming system can be effectively mapped on a random model
which is described in statistical terms. For this purpose we
generalize the simple trap model of Bouchaud and coworkers by
dividing the total system into M weakly interacting identical
subsystems, each being described in terms of a trap model. The
distribution of traps in this extended trap model (ETM) is fully
determined by the thermodynamics of the glass-former. The dynamics
is described by two adjustable parameters, one characterizing the
common energy level of the barriers, the other the strength of the
interaction. The comparison is performed for the standard binary
mixture Lennard-Jones system with 65 particles. The metabasins,
identified in our previous work, are chosen as traps. Comparing
molecular dynamics simulations of the Lennard-Jones system with
Monte Carlo calculations of the ETM allows one to determine the
adjustable parameters. Analysis of the first moment of the waiting
distribution yields an optimum agreement when choosing $M\approx
3$ subsystems. Comparison with the second moment of the waiting
time distribution, reflecting dynamic heterogeneities, indicates
that the sizes of the subsystems may fluctuate.

\end{abstract}

\pacs{0.8.15}
\maketitle

\section{\label{introduction}Introduction}

Understanding the properties of supercooled liquids and in
particular the nature of the glass transition is still a major
scientific challenge. An important concept to grasp the important
physics is related to the potential energy  landscape (PEL)
\cite{Goldstein:1969,Debenedetti:217,Wales:2003}. In this view the
system is regarded as a point in the high-dimensional
configuration space. At sufficiently low temperatures the system
is mainly characterized by the different minima ({\it inherent
structures}) of the PEL and their respective harmonic attraction
basins
\cite{Stillinger:1982,Sciortino:1999,Buchner:193,Mossa:284}. With
the advent of faster computers in recent years it became possible
to elucidate the properties of the minima and the saddles and thus
to relate the thermodynamic
\cite{Sciortino:1999,Buechner:1999,LaNave:265,Starr:140} as well
as the dynamic behavior of supercooled liquids
\cite{Scala:71,Keyes:94,Heuer:1997,Broderix:228,Schroder:210,Scala:376,Angelani:315,Wales:213,Denny:2003,Doliwa:2003a,Doliwa:2003b,Doliwa:2003c,Emilia04}
to the properties of the PEL. Most of the work has been devoted to
the study of the thermodynamics. From analyzing the PEL for
different densities it was possible, e.g., to extract the equation
of state over a very broad range of temperatures and pressures
\cite{Starr:140,Emilia:302}.

The Adam-Gibbs relation constitutes a relation between the
dynamics and the thermodynamics \cite{adam:39}. It expresses the
relaxation time in terms of the configurational entropy. This
relation seems to hold rather well for different systems
\cite{Sastry:198,Scala:71,Nicolas:411}. The theoretical relevance,
however, is not clear so far. Therefore it is still an open
question which properties of the PEL beyond the distribution of
inherent structures determine the dynamic behavior
\cite{Stillinger}.

\begin{figure}
  \includegraphics[width=8.0cm]{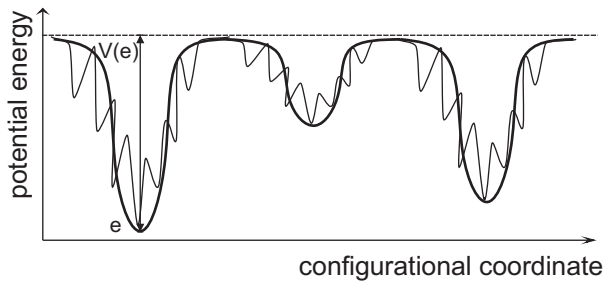}
  \caption{\label{fig1} Thick line: Sketch of the standard trap model. Thin line:
  The traps are identified
  as the metabasins and typically contain several inherent structures. }
\end{figure}

It has been attempted to fill this gap between thermodynamics and
dynamics by phenomenological models for which the system is
hopping between discrete states like for the random-energy model
 or the trap model \cite{Dyre95,Monthus:310,Odagaki95}. Here we
 are in particular interested in the trap model as discussed by
 Monthus and Bouchaud \cite{Monthus:310}.
 This model starts from a distribution $G(e)$ of traps with depth $e$. The
average waiting time in a trap of depth $e$ is given by $\tau(e,T)
= \tau_0 \exp(\beta V(e))$ where $V(e) = b - e$ and $b$ is the
common barrier level. After thermal excitation the system randomly
chooses a neighbor trap; see Fig.\ref{fig1} for a one-dimensional
sketch. The trap energies are randomly distributed. Of course, for
a real system one has to consider a distribution of traps in
high-dimensional space. At low temperatures the slow dynamics is
related to very long residence times in deep traps. Most work has
been devoted to exponentially distributed traps but different
distributions $G(e)$ can be chosen as well. In any event, $G(e)$
fully determines the thermodynamic properties and should reflect
the energy-distribution of inherent structures. A priori, it is
not clear whether the trap model is of major relevance for
describing the PEL of real glass-forming liquids. From a
conceptual point of view an important property of the trap model
is its non-topographic nature.  The arrangement of traps is purely
statistical and the dynamics is exclusively related to the
properties of the individual traps and not to possibly subtle
correlations among adjacent traps. We mention in passing, that in
alternative approaches the dynamics is rationalized without any
reference to thermodynamic properties \cite{Garrahan02}.

A very detailed analysis of the dynamics naturally involves the
properties of the barriers between the inherent structures as well
as the topology of the PEL. For a binary-mixture Lennard Jones
system (BMLJ)  trajectories, generated via molecular dynamics
simulations, have been analyzed in detail. It is immediately
visible that during very long times the system resides in some
fixed region of the configurational space. Thus the dynamics is
restricted to simple back and forth jumps between a finite number
of inherent structures \cite{Buchner:11,Denny:2003,Doliwa:2003b}.
It is useful to regard such a set of inherent structures as an
elementary state, denoted metabasin (MB)
\cite{Stillinger:333,Middleton:214,Doliwa:2003a,Aimorn:2003}. The
energy and configuration of a MB is taken from the inherent
structure of this MB with the lowest energy. Finally, the total
trajectory of the system in configuration space can be regarded as
a sequence of different MB, each characterized by an energy and a
waiting time. Actually, some work has been recently performed to
relate this configurational space picture to real space dynamics
\cite{Middleton:214,Reinisch:2004a,Reinisch:2004b,Vogel:2004,Trygubenko:2004}
and to explain the response to applied strain in glasses with the
help of the concept of MB \cite{Osborne04}.

Here we would like to remind the reader of a remarkable result,
presented in previous work \cite{Doliwa:2003b}. It turns out that
for all relevant temperatures analyzed so far,
\begin{equation}
\label{eqrw} \sum_{i=1}^N \langle (r_i(n) - r_i(0))^2 \rangle
\approx a^2 n.
\end{equation}
The left term denotes the mean square displacement after n
MB-transitions and $a^2$ is the temperature-independent average
distance between two adjacent MB in configuration space. Two
important consequences follow from Eq. \ref{eqrw}: (i) On the
level of MB the dynamics of the BMLJ system can be described as a
random walk in configuration space; (ii) The diffusion constant
can be expressed as \cite{Doliwa:2003b}
\begin{equation}
\label{diffusion}
 D(T) = a^2/6\langle \tau (T) \rangle.
\end{equation}
 Thus knowledge of the average
waiting time is sufficient to predict the macroscopic dynamics.
Actually, the simplicity of these results can serve as the a
posteriori justification for the introduction of the MB approach
because on the level of inherent structures no simple physical
picture can be formulated for the dynamics of the BMLJ system
\cite{Doliwa:2003b}. Recent work for a pure Lennard-Jones system
of significantly smaller size seems to indicate that there the
random-walk like dynamics already holds on the level of inherent
structures \cite{Keyes:335}.

Another remarkable result is the fact that a BMLJ system with
$N=65$ particles has no relevant finite size effects for the
dynamics in the accessible temperature range for computer
simulations\cite{Doliwa:404}. In particular, a system with $N=130$
particles behaves like an {\it independent} superposition of two
systems with $N=65$ particles. Therefore it is sufficient to study
the PEL of the BMLJ system with $N=65$ particles. Actually, when
choosing even smaller systems (e.g. $N=40$) significant
finite-size effects occur \cite{Buchner:193}.

Maybe the simplest disordered model, reproducing the observed
features, is the trap model as introduced above. Since the
distribution of traps is fully determined by the thermodynamics of
the system, the dynamics is basically governed by a single
parameter, namely the barrier height $b$.  In any event, if the
PEL can be mapped on a trap model, the MB have to be identified as
traps; see Fig.\ref{fig1}. Recent simulation work for the BMLJ
system directly shows in different ways that on a quantitative
level the predictions of the trap model are neither compatible
with  the distribution of waiting times \cite{Denny:2003} nor with
the activation energy $V(e)=b-e$ \cite{Doliwa:2003a}.

The scope of this work is twofold. First, we will use general
arguments to define an extended version of the trap model. It
fulfills some basic requirements which any model for glass-forming
systems should fulfill. Second, by detailed comparison with the
properties of the BMLJ system we elucidate to which level this
extended trap model will be able to explain the BMLJ dynamics.
From the remaining differences important information about the
nature of the PEL can be derived. Among other things the concept
of dynamic heterogeneities is related to properties of the PEL in
a detailed way. Finally, some possible applications of the present
results will be indicated.

\section{\label{model}The BMLJ system}

\subsection{Definition}

The BMLJ system is one of the standard glass-forming systems used
in computer simulations. It contains 80\% large and 20\% small
particles. We use the potential parameters as outlined in
Ref.\cite{Kob:203,Doliwa:2003a}. We have performed molecular
dynamic simulations for temperatures between $T = 0.4$ and $T = 1$
with system size $N=65$ and using periodic boundary conditions.
Here we present data for $T \ge 0.45$. All units are given in LJ
units. The mode-coupling temperature for this model is $T_c
\approx 0.45$ \cite{Goetze:1992,Doliwa:2003b}. As shown in
previous work this system size is large enough to reproduce the
macroscopic behavior of the BMLJ at least for $T \ge T_c$.

\subsection{Some previous results}

The most important quantity for the thermodynamics is the
effective density of MB, i.e. $G(e)$.
 In
previous work it turned out that to a very good approximation
\begin{equation}
G(e) \propto \exp(-(e-e_0)^2/2\sigma_0^2)
\end{equation}
with $e_0 = -279.2$ and $\sigma_0^2 = 9.05$ in Lennard-Jones units
\cite{Sciortino:1999,Buchner:193,Doliwa:2003a}. From entropy
arguments one can estimate that $G(e)$ has a low-energy cutoff
around $e = -306$. From $G(e)$ one can calculate the Boltzmann
probability $p(e,T)$ that at some randomly chosen time the system
visits a MB of energy $e$ at temperature $T$ via
\begin{equation}
\label{def_p} p(e,T) \propto G(e) \exp(-e/T).
\end{equation}
Note that $G(e)$ is slightly different from the real density of MB
due a small dependence of the local curvature of the different MB
on their energy $e$ \cite{Sciortino:1999,Buchner:193}. We just
mention in passing that basically the same results for the
thermodynamics are obtained if  inherent structures rather than MB
are analyzed.

\begin{figure}
  \includegraphics[width=8.0cm]{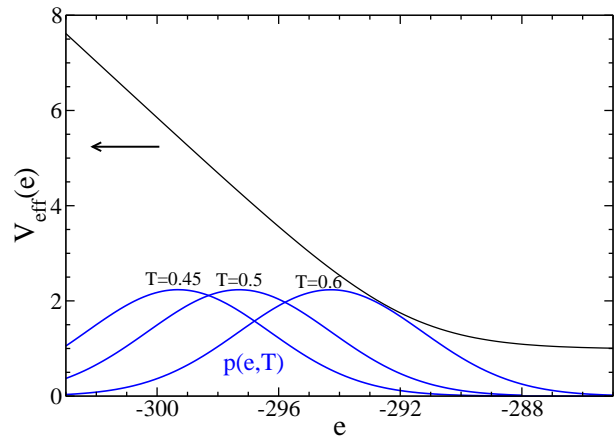}
  \caption{\label{figve} The function $V_{eff}(e)$ as obtained for the BMLJ system  together
  with the distribution $p(e,T)$ for three different temperatures.}
\end{figure}

In previous simulations we have analyzed in detail the waiting
time of the system in MB of energy $e$. Averaging over different
MB with energy $e$ we have found the relation
\begin{equation}
\label{taue} \langle \tau(e,T) \rangle = \tau_0(e) \exp(\beta
V_{eff}(e)),
\end{equation}
i.e. the average waiting time in a MB of energy $e$ shows simple
Arrhenius behavior. The function $V_{eff}(e)$ can be interpreted
as the average saddle height when leaving a MB of energy $e$. This
function is reproduced in Fig.\ref{figve}. Note that it was
possible to identify the saddles in the PEL which give rise to the
specific values of $V_{eff}(e)$. For $e < -295$ the function
$V_{eff}(e)$ is linear in $e$ with slope -0.55. In the high-energy
limit $V_{eff}(e)$ approaches a constant close to 1. This reflects
the fact that the dynamics in high-energy states resembles a
non-activated fluid-like dynamics. Using this choice of
$V_{eff}(e)$ it is possible to formally describe this fluid-like
dynamics as transitions between MB \cite{Doliwa:2003a}. A similar
limiting value is reported in \cite{Chowdhary04}. In the same
energy range the effective prefactor $\tau_0(e)$ increases by one
order of magnitude \cite{Doliwa:2003a}.

As shown in \cite{Doliwa:2003c} the dynamics is governed by
activated processes for $T \le 0.6$. Correspondingly, the average
waiting time is dominated by the escape from low-energy MB ($e \le
-295$). For temperatures below $T_c$ it becomes difficult to
obtain equilibrium data. Therefore we restrict our detailed
analysis to temperatures between 0.45 and 0.6 to compare the
dynamics of the BMLJ system with the predictions of the modelling.
In Fig.\ref{figve} we have also included the distributions
$p(e,T)$ in this temperature range, obtained from Eq.\ref{def_p}.
On the low-energy side the contribution stems from few events with
relatively large waiting times. Thus energy-dependent observables
for,  say, $e < -300$ possess rather large statistical errors. In
the detailed analysis of the subsequent sections we consider MB
with $e \in [-300,-295]$.

\section{Extended Trap Model: Definition and interpretation}

\subsection{General aspects of statistical models} How do properties of the PEL translate into dynamics?
According to the landscape paradigm the properties of
glass-forming systems are reflected by the properties of the
minima. As discussed above, we would use the MB rather than the
minima as the basic objects. In any event, one has an
exponentially large but finite number of states. In the most
detailed description the dynamics out of some state $i$ is
governed by two temperature dependent functions. First, the
probability function $r(...,i \rightarrow j)$ describes the
probability that the next state visited is state $j$. The dots
indicate that in principle this probability may also depend on the
history before entering state $i$. Second, the waiting time in
state $i$ is characterized by the distribution function of the
waiting times $f(\tau|...,i\rightarrow j)$ when going from state
$i$ to state $j$.

\begin{figure}
  \includegraphics[width=8.0cm]{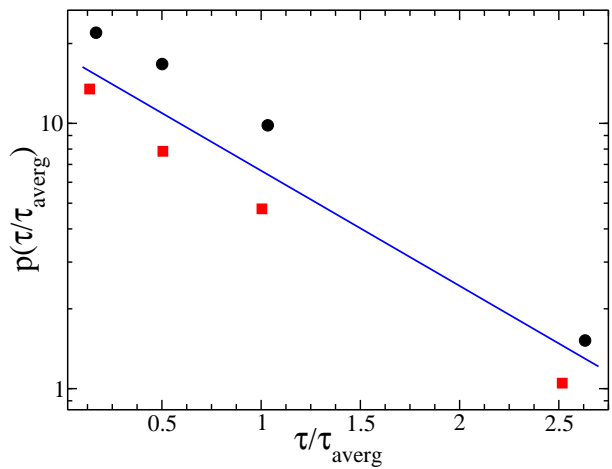}
  \caption{\label{figexp} The probability distribution of waiting times for two randomly chosen MB.
  The time-axis has been normalized by the respective average value $\tau_{averg}$. A perfect
  exponential waiting time distribution with slope -1 has been included. For better visibility the curves are shifted
  against each other. 60 escape events have been analyzed for each MB.}
  \end{figure}

Of particular interest are the low-energy states because they
determine the dynamics at low temperatures. As shown in
\cite{Doliwa:2003c} the escape out of these states is activated.
This implies that the dynamics can be very well described as
Markovian, i.e. the escape out of state $i$ does not depend on the
previous states. Furthermore the waiting time distribution is
close to exponential, i.e. characterized by a single average
waiting time. This is explicitly shown in Fig.\ref{figexp} for two
randomly chosen MB, both having energies close to $e = -302$. To
obtain one set of data, we have performed repeated simulations
from the same initial configuration, using different initial
velocities. Performing our standard MB analysis we have identified
for every run  the time when the system is leaving the MB.

As a consequence, the dynamics is fully characterized by the
probability function $r(i \rightarrow j)$ and the average waiting
time of state $i$. Alternatively, the information can be expressed
by the transition rates $\Gamma(i\rightarrow j)$. On this level,
the total dynamics can be formulated as a set of rate equations.
The rates can in principle be determined from simulations.
Actually, for very small systems where most relevant states and
corresponding saddles can be explicitly determined, rate equations
have been already studied \cite{Doye:194}. We mention in passing
that in this framework equilibrium as well as non-equilibrium
phenomena are fully characterized.

This approach has to fail for an exponentially large number of
relevant minima. Rather one has to resort to some statistical
approach. This involves two steps. First, one has to characterize
each state $i$ by some relevant internal parameters
$x_{i,1},x_{i,2},...$ .  As a natural choice energy may be taken
as one of these parameters. Then a direct connection to the
thermodynamics becomes possible. Second, one has to define the
transition rates $\Gamma_T(i\rightarrow j)$. In the trap model,
$e$ is the only internal parameter and $\Gamma_T(i\rightarrow j)$
does not depend on $j$. For a good statistical model it would be
possible to reproduce any time-dependence of the energy for any
temperature schedule, i.e. also in non-equilibrium situations.
Reproducing observables like the alpha-relaxation time or the
diffusion constant requires additional information about the
real-space nature of the different states. Fortunately, for the
BMLJ system(and also for BKS-SiO$_2$ (unpublished results)), Eq.
\ref{eqrw}  expresses a very simple relation between the
configuration space dynamics and the temperature dependence of the
diffusion constant.

Searching for a very simple statistical model one may naturally
end up with a trap model because it only involves the energy as
the only internal parameter. Beyond its simplicity it has the
advantage that of automatically reproducing the random-walk like
nature of dynamics (Eq.\ref{eqrw}). However, as mentioned in the
Introduction major discrepancies between the BMLJ data and the
predictions of the trap model have been reported. In the following
we will formulate an extension of the trap model by including more
than one internal parameter. Then it will be possible to reproduce
several non-trivial aspects of the BMLJ dynamics.

\subsection{Relaxation in a collection of elementary trap systems}

\begin{figure}
  \includegraphics[width=8cm]{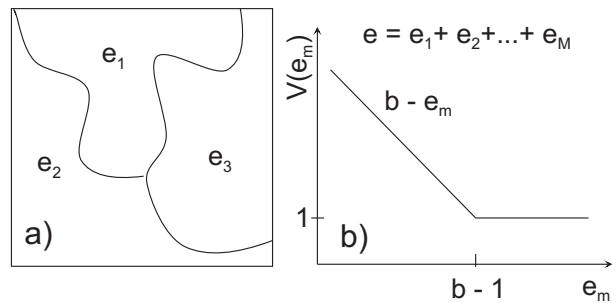}
  \caption{\label{fig2}(a) The division of the total system in M independent subsystems (here: M=3).
  Of course, the borders are expected to fluctuate with time. (b) Every subsystem
  is described by a trap-model, which is characterized by the function $V(e)$. }
\end{figure}

It is well accepted that for supercooled liquids there exists a
finite dynamic correlation length, sometimes denoted length scale
of dynamic heterogeneities
\cite{Doliwa:spatial,Donati:94,Perera:164}. It can even be
measured experimentally \cite{Tracht,Reinsberg01}. At a given
temperature, on average $P$ particles ($P$ may be somewhat
temperature dependent) relax in a collective fashion. Stated
differently, relaxation processes are restricted to small regions
of the system, containing on average $P$ particles. Depending on
the exact definition of the length scale of dynamic
heterogeneities one expects values between $10$ and $100$
particles. Thus one may be tempted to divide the total system into
$M = N/P$ subsystems which to first approximation behave
independently \cite{Doliwa:404,Chowdhary04}.  Of course, the
division into $M$ subsystems should not be taken too literally.
Rather these border lines may fluctuate with time (see also
below). Guided by the random-walk nature of the dynamics in
configuration space we approximate every individual subsystem by a
trap model; see Fig.\ref{fig2}a. The resulting model is termed
{\it extended trap model} (ETM).  Then the total energy $e$ may be
written as $e = \sum_{m=1}^M e_m$. A simple trap model would
correspond to $M=1$. However, in this case one would have expected
that $V_{eff}(e)$ has a slope of -1 for low $e$ in contrast to the
real behavior; see Fig.\ref{figve}. Thus it is already clear that
a mapping on a simple trap model is not possible and that, if at
all, $M > 1$ has to be chosen for a BMLJ system with N=65
particles\cite{Denny:2003,Doliwa:2003a}.

In order to reproduce the thermodynamics of the BMLJ system the
distribution of traps $G(e)$ in each subsystem has to be chosen as
\begin{equation}
G_M(e_m) \propto
\exp\left(-\frac{(e_m-(e_0/M))^2}{2(\sigma_0^2/M)}\right ).
\end{equation}
The dynamics is expected to strongly depend on the length scale of
dynamic heterogeneities, i.e. on the value of $M$. In what
follows, we will present results for different values of $M$ and
identify the optimum choice.

As discussed above, the most relevant parameter of any trap model
is the common barrier level $b$. For low energies this gives rise
to the choice $V(e) = b-e$ as the barrier height to leave a trap
(i.e. MB) of energy $e$. To cope with the high-energy limit of
$V_{eff}(e)$, discussed above, we choose $V(e) = 1.0$ for high
$e$. The precise value is irrelevant for the results in the
relevant temperature range. In total, we have $V(e) = {\rm
max}(b-e,1)$. This function is sketched in Fig.\ref{fig2}b. The
average waiting time in a trap of energy $e$ is given by
\begin{equation}
\label{eqtau} \tau(e,T) = \tau_0 \exp(\beta V(e)).
\end{equation}
 The value of
$\tau_0$ is a trivial scaling factor for the overall time scale
and can be simply adjusted.

\subsection{Dynamic interaction between subsystems}

The ETM, introduced so far, starts from a collection of $M$
totally independent subsystems. However, one should rather expect
that a local relaxation process may somewhat influence the state
of the adjacent subsystems. Thus, some kind of {\it dynamic}
interaction should be included, i.e. a modification of the state
of an adjacent subsystem as a result of a relaxation process. In
real space this modification might correspond to a minor shift of
particles in the adjacent subsystems. This shift may somewhat
change the probability for a relaxation process in these
subsystems. Actually, already in the original paper of Monthus and
Bouchaud their master equation for the energy probability
distribution has been supplemented by an energy diffusion term
which exactly takes into account the effect of relaxation
processes on adjacent particles \cite{Monthus:310}.

Here we introduce a simple ingredient of the ETM which reflects
the relevant physics of the dynamic interaction. In particular we
will take care that the thermodynamics of the total system is not
modified. This condition is non-trivial because any variation of
states may directly influence the probability distribution of
states and thus change quantities like the average energy.

For the total system one may define the distribution
$\varphi(e,T)$ via
\begin{equation}
\label{phi_def} \varphi(e,T) \propto p(e,T) /\langle \tau(e,T)
\rangle.
\end{equation}
It denotes the probability that after or before a transition a MB
has the energy $e$. Thus, generation of MB according to the
$\varphi(e)$-distribution (for convenience we suppress the
dependence on T) yields the correct statistics of MB and thus the
correct thermodynamics. Conceptually, the distribution
$\varphi(e)$  is close to the distribution of MB, i.e.  $G(e)$.
Some minor (temperature dependent) deviations are present because
in the present version of the trap model the relation $V(e) = b -
e$ is not valid for the high-energy traps with $e > b-1$.

In what follows we generalize Eq.\ref{phi_def} to the case $M >
1$. First we consider the case $M=2$ but generalizations are
straightforward. Let the subsystem 1 be in state $e_1$ and the
subsystem 2 in state $e_2$. In the ETM the transition of the total
system is achieved by a transition of one subensemble. Then the
rate $1/\tau(e_1,e_2)$ for a transition of the system is given by
the sum of the individual rates, i.e.

\begin{equation}
\label{tau2e}
 1/\tau(e_1,e_2) = 1/\tau(e_1) + 1/\tau(e_2).
\end{equation}

Furthermore due to the {\it independence} of the two subsystems
the combined Boltzmann probability $p(e_1,e_2)$ is identical to
the individual Boltzmann probabilities, i.e.
\begin{equation}
p(e_1,e_2) = p(e_1) p(e_2).
\end{equation}

Finally, the probability that after (or before) a transition of
the total system one finds energies $e_1$ and $e_2$ is given by

\begin{eqnarray}
\varphi(e_1,e_2) & \propto & p(e_1,e_2)/\tau(e_1,e_2) \nonumber \\
&\propto&
p(e_1) p(e_2)(1/\tau(e_1) + 1/\tau(e_2)) \nonumber \\
& \propto & \varphi(e_1) \tau(e_1) \varphi(e_2) \tau(e_2) (1/\tau(e_1) + 1/\tau(e_2)) \nonumber \\
& \propto & \varphi(e_1) \varphi(e_2) ( \tau(e_1) + \tau(e_2)).
\end{eqnarray}

Due to the enormous dynamic heterogeneities (see below) the
waiting times are distributed over many orders of magnitude. Thus,
at a randomly chosen time one will typically find both subsystems
in traps of very different waiting times $\tau_1$ and $\tau_2$.
Let us assume that a jump process occurred in subsystem 1. Then it
is very likely that the state before the jump was characterized by
$\tau_1 \ll \tau_2$. Therefore one can write
\begin{equation}
\label{phi_p} \varphi(e_1,e_2) \propto \varphi(e_1) \varphi(e_2)
\tau_2 \propto \varphi(e_1) p(e_2).
\end{equation}

Now we may define the dynamic interaction. It is modelled such
that after a jump in one subsystem with probability $q$ a new trap
is selected in the other subsystem. According to Eq.\ref{phi_p}
the energy of the new trap in every subsystem has to be selected
from the probability distribution $p(e_2)$. This implies that the
thermodynamics is not modified by the dynamic interaction.
Furthermore, the principle of causality is directly implemented.
 Of course, in practice one would expect that
every jump induces a minor rearrangement so that after $1/q$ jumps
of the fast subsystem a total rearrangement with respect to the
p-distribution has occurred. This gradual process, however, is
very well approximated by the random process, introduced above.

Actually, for the low-energy limit, i.e. for  $\tau(e_2) \propto
\exp(\beta(b - e_2))$, there exists a more intuitive
rationalization of the factor $p(e_2)$ in  Eq.\ref{phi_p}. It is
reasonable to assume that the probability for subsystem 2 to
change its energy from $e_2^\prime$ to $e_2$ will be proportional
to $\varphi(e_2)$ and the Boltzmann factor $\exp(\beta(e_2^\prime
- e_2))$. Therefore the probability to be in state $e_2$ is
proportional to $\varphi(e_2)\exp(\beta(e_2^\prime - e_2) \propto
\varphi(e_2)\exp(-\beta e_2) \propto \varphi(e_2)\tau(e_2) \propto
p(e_2)$. Of course, the original derivation of Eq.\ref{phi_p} is
more general.

For general $M$ one obtains the relation
\begin{equation}
\label{phi_p_gen} \varphi(e_1,...,e_M) \propto \varphi(e_1) p(e_2)
\cdot \cdot \cdot p(e_M)
\end{equation}
if a jump occurred in the first subsystem. Here we modify all
subsystems $m=2,...,M$ with probability $q$ according to the
respective Boltzmann distribution $p(e_m)$. Evidently, in reality
the dynamic interaction between subsystems may be somewhat more
complicated. It will turn out, however, that this simple choice
already captures the main effect of the interaction.

\subsection{How to compare the ETM dynamics with the BMLJ dynamics?}

So far we have introduced the ETM with three parameters $M$, $b$,
and $q$. Every state is characterized by the internal parameters
$e_1,...,e_M$. The time evolution of the system according to the
ETM is generated in a straightforward way. After a relaxation
process in one of the $M$ subsystems a new energy is selected for
this subsystem according to the {\it $\varphi$-distribution},
defined in Eq.\ref{phi_def}. Furthermore with probability $q$ a
new energy is selected for the other $M-1$ subsystems according to
the $p-distribution$, defined in Eq.\ref{def_p}. The {\it total}
energy of all subsystems is denoted $e^i$ (the upper index counts
the number of transitions of the total system) . In the next step
for every subsystem a waiting time $\tau_{act}(e_m)$ is
calculated. It is drawn from an exponential distribution with the
average value $\tau(e_m,T)$; see Eq.\ref{eqtau}. Then the
subsystem with the shortest waiting time $\tau_{act,min}$ is
selected to perform a relaxation process and the time proceeds by
exactly $\tau^i \equiv \tau_{act,min}$. This procedure is then
repeated and in the next step $e^{i+1}$ and $\tau^{i+1}$ are
obtained etc. The output is a sequence of waiting times
$\{\tau^i\}$ and corresponding energies $\{e^i\}$. Exactly these
two sequences can be also obtained from analysis of the
equilibrium BMLJ dynamics.

Both series together characterize in a detailed way how the system
is exploring the PEL. They are to a large extent characterized by
(i) the first and second moments of their respective distribution
functions and (ii) by their correlation functions
\begin{equation}
f_\mu(n) \equiv \langle \mu_{n} \mu_{0} - \langle \mu \rangle^2
\rangle / \langle (\mu - \langle \mu \rangle)^2 \rangle.
\end{equation}
The variable $n$ counts the number of MB transitions and $\mu \in
\{\tau,e\}$. The denominator guarantees that $f_{\mu}(0) = 1$.
Basically, the $f_\mu(n)$ express to which degree the waiting
times and the MB energies are correlated after $n$ MB transitions.
Since the trap distribution in the ETM is chosen in order to
reproduce the thermodynamics of the BMLJ system, the distribution
of energies is naturally recovered. Thus the observables $\langle
\tau\rangle$, $\langle \tau^2 \rangle, f_{\tau}(n)$ and $f_{e}(n)$
may be taken for comparison. If the ETM is an appropriate model to
characterize the PEL it should be possible to choose the model
parameters $M,b$ and $q$ such that the BMLJ results of these
observables are recovered for the relevant temperature regime
between 0.45 and 0.6.

It is important to clarify the influence of the dynamic
interaction parameter $q$ on the series of waiting times and
energies, respectively. Since our procedure does not change the
distribution $\varphi(e_1,e_2)$ (here for M=2) and since the
waiting times are directly related to the energies, the
distributions of waiting times and energies do not depend on $q$.
In contrast, due to the additional variation of energies (and thus
waiting times) one may expect that the decay of $f_\mu(n)$ becomes
faster with increasing $q$. Thus, except for comparison of
$f_\mu(n)$ we may choose $q=0$.

\section{Results of the mapping}

So far we have developed a minimum model for the PEL of a
glass-forming system based on the trap model as the elementary
unit. Now we compare the resulting ETM with the BMLJ system. As
already mentioned this comparison involves the time series of
waiting times as well as the time series of MB energies.

\subsection{Time series of waiting times}

\begin{figure}
  \includegraphics[width=8cm]{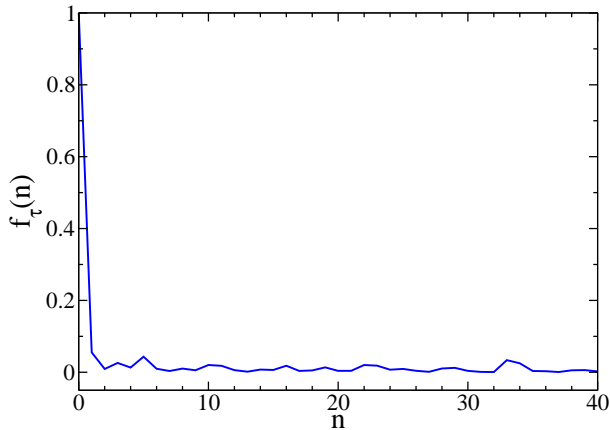}
  \caption{\label{fig3}The correlation function of waiting times in dependence of
  the number of MB transitions for the BMLJ system at $T=0.5$.}
\end{figure}

First, we analyse the autocorrelation function $f_\tau(n)$. The
result for the BMLJ system is shown in Fig.\ref{fig3}.
Interestingly, it turns out that already after one step the new
waiting time is basically uncorrelated to the old waiting time.
This observation can be easily rationalized in terms of the ETM.
From Eq. \ref{phi_p} (or Eq. \ref{phi_p_gen} for general $M$) it
follows that the new state of subsystem
 1 will be drawn according to the distribution
$\varphi(e_1)$. Since $\varphi(e_1)$ is peaked at much higher
energies than $p(e_2)$ it is very likely that also the next jump
occurs in subsystem 1 (see below for a closer analysis of this
effect). Since, however, two successive energies in subsystem 1
are uncorrelated (by definition) the same holds for the two
successive waiting times for this subsystem. Since these waiting
times determine
 the waiting time of the total system (because
subsystem 2 is basically immobile) one may conclude that in
general two successive waiting times of the total system are to a
very good approximation uncorrelated. Of course, it can be checked
explicitly (data not shown). This non-trivial prediction fully
agrees with the MD data for the BMLJ system.

Now we discuss the first and the second moment of the waiting time
distribution $f(\tau)$ which, of course, strongly depends on
temperature. As discussed above we may set $q=0$ for the
comparison with the BMLJ system. In a first step we determine the
respective values of $b$ (for different values of $M$) from the
condition that the temperature dependence of $\langle \tau
\rangle$ should agree as well as possible with the BMLJ data.

There are two different ways to estimate the value of $\langle
\tau \rangle$. On the one hand, one can calculate the average of
all waiting times, observed during a simulation run of time
$t_{sim}$. On the other hand, one may determine the number of
transitions $n$ and take $t_{sim}/n$ as the average waiting time.
As shown in Appendix I the second choice can be applied for any
simulation time $t_{sim}$. In contrast, the first choice is only
applicable for very large $t_{sim}$. Therefore we prefer to use
the second choice.

\begin{figure}
  \includegraphics[width=8.0cm]{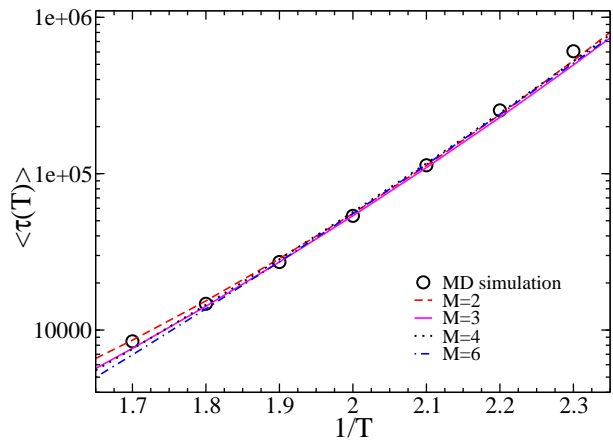}
  \caption{\label{fig4}The average waiting time of MD simulations as compared
  with the predictions of the ETM with different values of $M$. The resulting
  values of $b$ and $\tau_0$ for the different $M$ are given in Tab.1.}
\end{figure}

In Fig.\ref{fig4} we show the temperature dependence of the
average waiting time $\langle \tau(T) \rangle$ as obtained from
the MD simulations.  These data are used to determine an optimum
value of the common barrier level $b$ for the different choices of
$M$. The resulting predictions for the ETM are also included in
Fig.\ref{fig4}. The values for $b$, obtained from this fitting,
are listed in Tab.1. The agreement with the BMLJ data is not
perfect but this single fitting parameter $b$ is enough to
reproduce the temperature dependence in the relevant temperature
range between 0.45 and 0.6 in a semi-quantitative way. By
increasing the value of $b$ it would have been possible to improve
the agreement for low temperatures whereas the deviations at
higher temperatures would have become stronger. In any event, the
temperature dependence of the ETM for the different values of $M$
looks very similar. Deviations at higher temperatures are expected
because in this regime flow-like processes become more important
which, of course, cannot be fully captured by the ETM.

 \begin{table}
  \begin{center}
  \begin{tabular}[t]{|c|c|c|}\hline
       \multicolumn{1}{|c|}{\parbox[t]{1.0cm}{M}}&
       \multicolumn{1}{c|}{\parbox[t]{1.0cm}{b}} &
        \multicolumn{1}{c|}{\parbox[t]{1.0cm}{$\tau_0$}}\tabularnewline[1mm]
       \hline
       2   & -143.1 & 160 \tabularnewline[1mm] \hline
      3  & -93.3 & 43 \tabularnewline[1mm] \hline
       4  & -68.3 & 14\tabularnewline[1mm] \hline
       6   & -43.4 & 4 \tabularnewline[1mm] \hline
       \end{tabular}
  \caption{\label{tabsummary}The values of the barrier level $b$ and the prefactor $\tau_0$, obtained
  from the fitting as discussed in Fig.\ref{fig4}.}
  \end{center}
\end{table}

\begin{figure}
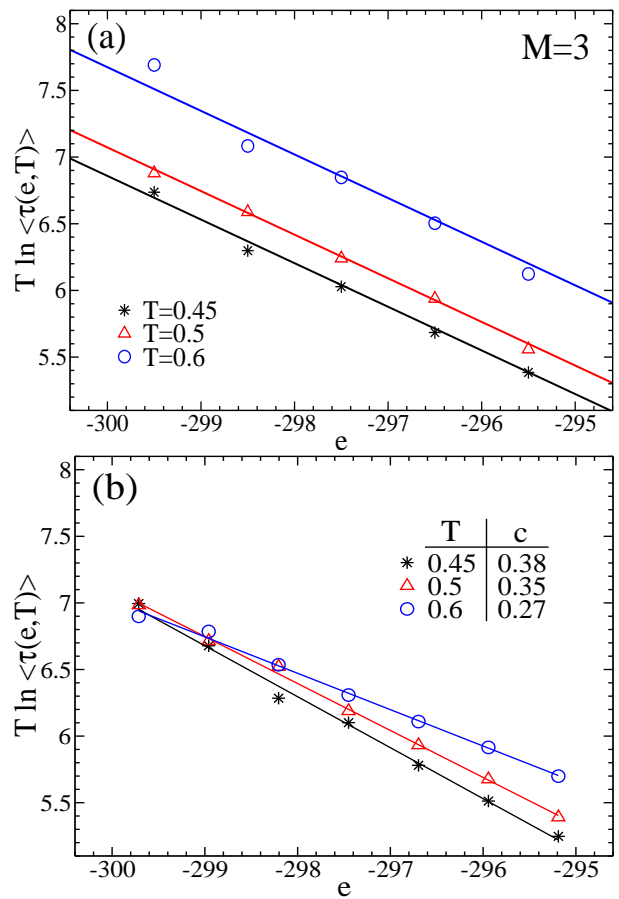

  \includegraphics[width=8.0cm,height=5.9cm]{Trapfig7a.eps}
  \includegraphics[width=8.0cm,height=5.9cm]{Trapfig7b.eps}
  \caption{\label{fig5} The quantity $ T \ln \langle \tau(e,T) \rangle$ for different
energies and temperatures in the energy range [-300,-295]. a)
Results for the ETM for the example $M=3$. The resulting slopes
for all relevant values of $M$ are listed in Tab.2. b) Results for
the BMLJ. The resulting temperature-dependent slopes are included
in the Figure.
 }
\end{figure}

So far we have analyzed the average waiting time. At a given
temperature it has contributions from MB with very different
energies. Therefore it is more informative to analyze the waiting
time in dependence of temperature {\it and} energy. As expressed
by Eq. \ref{taue} there is a significant dependence of the average
waiting time on the MB energy in the BMLJ system.  For this
purpose we consider the energy-dependence of the quantity $T
\ln\langle \tau(e,T) \rangle$. For a simple trap model (see Eq.
\ref{eqtau}) it is linear with slope -1. What is the result for
the ETM? The result for $M=3$ is shown in Fig.\ref{fig5}a for the
energy interval [-300,-295]. Interestingly, for all temperatures
(T=0.45;T=0.5;T=0.6) a constant slope is observed which
furthermore is identical for the three curves. For the energy
interval $[-300,-295]$ the data for the different temperatures can
thus be consistently described by
\begin{equation}
\label{defc}
 T \ln \langle \tau(e,T)\rangle \approx  const(T) - c e
\end{equation}
with $c = 0.33$. We have repeated the same analysis for the
different values of $M$. One observes the same behavior except for
different values of $c$; see Tab.2.

Actually, the value of $c$ can be also estimated from general
arguments which have been already partly presented in
\cite{Doliwa:2003a}. For $M$ subsystems and individual energies
$(e_1,...,e_M)$ in the limit of low temperatures the activation
energy is given by min$[V(e_1),...V(e_M)] = V(e_{max})$ where
$e_{max}$ is the largest value of the $\{e_m\}$. To calculate the
apparent activation energy $V_{eff}(e)$ for fixed total energy
$e$, one has to average this expression over different
realizations of the $e_m$ under the constraint $e = e_1 + ... +
e_M$. As a first approximation, however, one may take
$e_{max}\approx e/M + \Delta e$. The second term accounts for the
deviation of $e_{max}$ from the average value $e/M$. It can be
expected that $\Delta e$ only weakly depends on $e$. Then one has
\begin{equation}
V_{eff}(e) = V(e_{max}) = V(e/M + \Delta e) \approx const - (1/M)e
\end{equation}
Thus one expects that the slope $c$ is close to $1/M$ which is in
excellent agreement with the numerical results in Tab.2.

In analogy to the ETM we have calculated the quantity $T
\ln\langle \tau(e,T) \rangle$ for the BMLJ system for the same
temperatures. The results are displayed in Fig.\ref{fig5}b.
Interestingly, the data also display a linear energy-dependence.
Therefore it is possible to directly extract the value of $M$ if
the BMLJ data are interpreted in terms of a ETM. For the two
lowest temperatures the resulting slope is closest to $M=3$. For
reasons of simplicity we also choose $M=3$ for the $T=0.6$.  The
weak temperature-dependence of the slope will be further discussed
below.  We note in passing that for a monatomic Lennard-Jones
system of 32 particles the average barrier height linearly depends
on energy with slope one. This would correspond to M=1 and is thus
qualitatively compatible with the present result for a larger (and
somewhat different) system \cite{Chowdhary04}.

\begin{table}
  \begin{center}
  \begin{tabular}[t]{|c|c|}\hline
       \multicolumn{1}{|c|}{\parbox[t]{1.0cm}{M}}&
       \multicolumn{1}{c|}{\parbox[t]{1.0cm}{c}}\tabularnewline[1mm]
       \hline
       2   & 0.44 \tabularnewline[1mm] \hline
       3  & 0.33 \tabularnewline[1mm] \hline
       4  & 0.25 \tabularnewline[1mm] \hline
       6   & 0.17\tabularnewline[1mm] \hline
       \end{tabular}
  \caption{\label{tabsummary} The slope $c$ obtained from analyzing the energy
dependence of $T \ln \langle \tau(e,T) \rangle$ in
Fig.\ref{fig5}.}
  \end{center}
\end{table}

In the next step we want to elucidate the nature of the
fluctuations. A natural choice would be the quantity
\begin{equation}
\label{defvar} S(T) \equiv \frac{\langle \tau^2 \rangle-\langle
\tau \rangle^2}{\langle \tau \rangle^2}
\end{equation}
including the first and the second moment of the waiting time
distribution $f(\tau)$. It can be regarded as a measure for the
normalized variance of the waiting time distribution. Around $T_c$
one approximately finds $f(\tau) \propto \tau^{-2}$ for large
$\tau$. Without a cutoff of $f(\tau)$ at long times the variance
would diverge for this distribution. This implies that the
variance is extremely sensitive on the cutoff-behavior of
$f(\tau)$. For temperatures around $T_c$ this cutoff can be
predicted from Eq.\ref{eqtau} by setting $\epsilon = -306$ as an
estimate for the low-energy cutoff of the energy landscape. It
turns out $\tau_{max} \approx 10^5 \langle \tau \rangle$ which
corresponds to roughly $10^{10}$ MD steps. Only for simulation
times much longer than $\tau_{max}$ and thus many orders of
magnitude longer than possible by present computer technology the
variance of the observed waiting time distribution can be
determined from simulations around $T_c$. This will be explicitly
demonstrated in the next Section.

\begin{figure}
  \includegraphics[width=8.0cm]{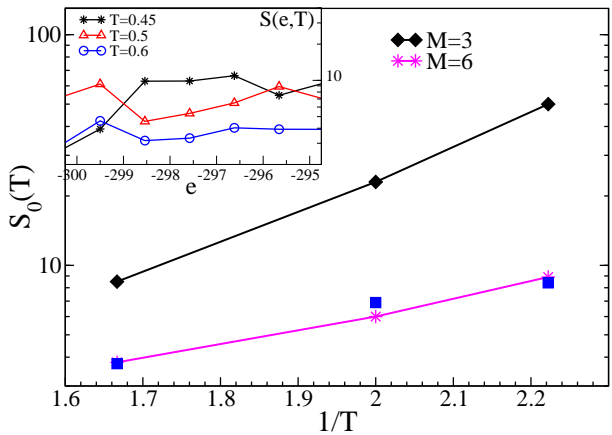}
  \caption{\label{fig6} Inset: Energy-dependence of the normalized variance
$S(e,T)$ for different temperatures for the BMLJ system. Main
panel: Comparison of the respective average values $S_0(T)$ with
the predictions of the ETM for $M=3$ and
 $M=6$.}
\end{figure}

Evidently, the very long waiting times result from MB with very
low energies. Therefore it may be very instructive to analyze the
fluctuations of waiting times of MB, restricted to energy $e$.
This restricted quantity is denoted as $S(e,T)$. For very low
energies one may expect that due to the extremely small number of
contributing MB the estimated variance will be smaller than the
true variance. Only for higher energies $S(e,T)$ can be estimated
without systematic errors from finite simulation times. We
restrict ourselves again to the interval $[-300,-295]$. $S(e,T)$
is expected to show a sensitive dependence on the nature of the
ETM. If, for example, the total BMLJ system could be described as
a single trap system ($M=1$), MB with energy $e$ would be
described by a single relaxation rate, yielding $S(e) = 1$. For $M
> 1$ the total energy can be decomposed in different ways,
resulting in different values of $\tau(e_1,...,e_M)$ for fixed
total energy $e_1 + ...+ e_M$; see Eq.\ref{tau2e}. The resulting
distribution is a sum of different exponentially distributed
functions, giving rise to $S(e)> 1$. The data for the BMLJ system
for $T = 0.45; 0.5; 0.6$ are shown in the inset of Fig.\ref{fig6}
for the interval $e \in [-300,-295]$. The average value of
$S(e,T)$ in this interval is denoted $S_0(T)$. It is plotted in
Fig.\ref{fig6} for the three temperatures. Furthermore we have
added the predictions for $S_0(T)$ of the ETM for $M=3$ and $M=6$.
It turns out that the data are inconsistent with the predictions
of the ETM for $M=3$. Rather a good agreement with $M=6$ is
suggested. As will be discussed further below a very natural
mechanism may exist which reduces the size of the fluctuations of
waiting times for a real system as compared to the predictions of
the ETM. Therefore we propose that $M\approx 3$ as obtained from
the analysis of $\langle \tau(e,T)\rangle$ is indeed the relevant
value for the ETM.

\subsection{Sequence of energies}

\begin{figure}
  \includegraphics[width=8.0cm]{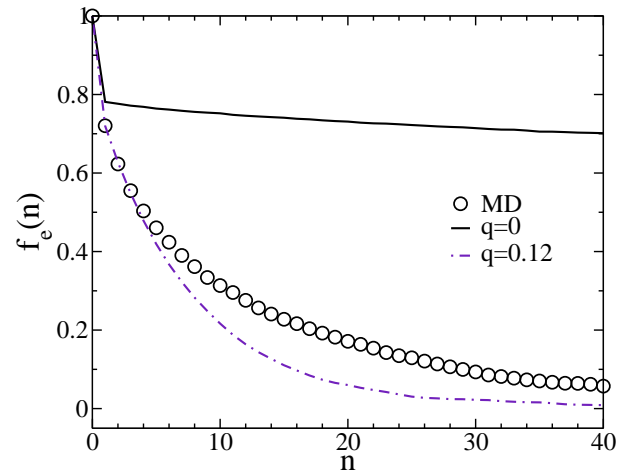}
  \caption{\label{fig7} Decay of the energy-autocorrelation function
in dependence of the number of MB transitions at T=0.5. Shown are
the MD data for the BMLJ system (circles) and the predictions with
($q=0.12)$ and without ($q=0.0)$ dynamic coupling for the case
$M=3$. }
\end{figure}

In a last step we analyse the sequence of MB energies. Since the
energy distribution of the real system is naturally  recovered by
the trap model (see above) the  non-trivial aspect is related to
possible correlations of  successive MB energies as measured by
the autocorrelation function $f_e(n)$. The function $f_e(n)$,
obtained for the BMLJ system, is shown in Fig.\ref{fig7}.  Since
for $T=0.5$ one has $\tau_\alpha \approx 20 \langle \tau \rangle$
($\tau_\alpha$ determined from the incoherent scattering function)
the energy correlation decays on the time scale of the
$\alpha$-relaxation.

Dramatic differences are visible when comparing  $f_e(n)$ with the
predictions of the ETM (M=3)without including the dynamic
interaction, i.e. $q=0$. Whereas the energies for the BMLJ system
significantly decorrelate after 40 steps, basically no
decorrelation is seen for the ETM. This can be easily
rationalized. In the typical scenario, discussed above, very often
there exists one subsystem which is much faster than the other two
subsystems and, correspondingly, has a relatively high energy.
This fast subsystem jumps and, on average, will acquire a new
waiting time which is still much faster than the waiting time of
the other two subsystems. Thus the energy of the slower systems
will remain unchanged even after a large number of transitions.
This automatically implies strong correlations for successive
total energies of the ETM.

In contrast, a finite dynamic interaction results in additional
reorganisation processes of the slow subsystems, thereby changing
their energy according to the p-distribution. Although $p(e,T)$ is
shifted to lower energies as compared to $\varphi(e,T)$ there will
be the chance to acquire a relatively high energy after a few
transitions of the fast subsystem and start to contribute to the
relaxation. Therefore one can expect that a finite $q$ will give
rise to a considerable decorrelation of energy. This can be seen
in Fig.\ref{fig7} where data for $q=0.12$ are included. With this
value the initial part of the decorrelation can be reproduced.
Taking the strong dependence of $f_e(n)$ on the value of $q$ into
account one may estimate the uncertainty of this value by $\pm
0.02$. Actually, a similar analysis for the two other temperatures
yield $q$-values in the same interval although the data suggest a
minor temperature-dependence ($q(T=0.6) \approx 0.14; q(T=0.45)
\approx 0.11$).

 Of
course, this mechanism of dynamic interaction is far too simple to
explain the complex coupling between different relaxation modes in
supercooled liquids. In particular, it turns out that finally the
decorrelation, generated by the dynamic interaction, is too
strong. From a physical point of view one would expect that states
with lower energies are less sensitive to fluctuations in the
neighborhood than states with somewhat higher energies. This
implies a dependence of the strength of the dynamic interaction on
energy. Qualitatively, this property would explain the remaining
deviations in Fig.\ref{fig7}. Unfortunately, when introducing an
energy-dependent $q$-value one would be forced to modify the
simple rule based on Eq.\ref{phi_p_gen} in order to obtain the
same thermodynamic properties. This is beyond the scope of the
present work and would not yield any relevant new insight into the
nature of the PEL of the BMLJ system.

\section{Discussion}

\subsection{Length scales and finite-size effects}

Analysing the time series of waiting times the BMLJ system with 65
particles behaves to a first approximation as if it were a
superposition of $M \approx 3$ independent subsystems. Each
subsystem contains on average 20-25 particles. However, performing
MD simulations with this very small number of particles (and
periodic boundary conditions) one would observe strong finite size
effects. Thus, it is not possible to perform simulations on the
level of the elementary units of the ETM. In this sense it is
impossible to see the elementary trap system directly. Rather the
properties of the individual subsystems, mainly expressed by the
value of the barrier $b$, have to be extracted from the
superposition of at least 2-3 subsystems, corresponding to roughly
65 particles in real space. Why does a PEL of a BMLJ system with
only 20 particles show significant finite size effects? First, the
presence of dynamic interaction directly shows that the coupling
among adjacent regions of the supercooled liquid is essential for
the dynamics. Second, one can imagine that the real-space pattern
of those particles which move together during one MB transition
may be shaped in an string-like manner
\cite{Donati:94,Perera:164}. Therefore the minimum size system
without major finite size effects has to be larger than the
elementary subsystem of the ETM. Otherwise the real-space patterns
cannot be formed appropriately.

Of course, one could repeat the analysis also for larger systems.
Evidently, larger values of $M$ would have been found. In any
event, since a BMLJ system with $N=130$ particles behaves like a
superposition of two systems with $N=65$ particles one would end
up with a ETM which is a trivial extension of the present system.

\subsection{How to rationalize the size of waiting time fluctuations at constant energy?}

The only observable which did not agree with the predictions of
the ETM was the second moment of the waiting time distribution.
The fluctuations at constant energy as expressed by $S_0(T)$ are
smaller than expected for $M=3$. Is there a simple physical
picture which may reconcile these results?

Recently a detailed real-space analysis of the nature of the MB
transition has been performed for the BMLJ system
\cite{Vogel:2004}. In particular the number of particles,
participating at such a transition, has been studied. This number
can be quantified by the participation number
\begin{equation}
\label{defz}
 z = \sum_i \frac{dr_i}{dR}.
\end{equation}
Here the sum is over all large particles of the binary
Lennard-Jones system. $dr_i$ is the displacement of the i-th
particle and $dR^2$ is the total squared displacement of all large
particles. In case that only one particle is moving one has $z=1$,
in case of identical movement of $n$ particles one gets $z=n$.
 It
turns out that for different MB transitions the participation
number $z$ can have very different values. The resulting
distribution $p(z)$ has significant contributions between $z
\approx 8$ and $z \approx 25$. Since the absolute numbers somewhat
depend on the definition of $z$ \cite{Reinisch:2004b} we mainly
stress the width of $p(z)$ rather than the absolute numbers.

\begin{figure}
  \includegraphics[width=8.0cm]{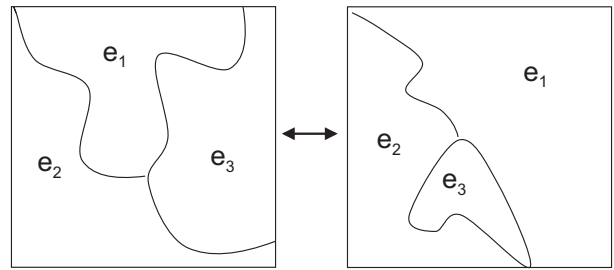}
  \caption{\label{fig9}Sketch of the size fluctuations of a ETM with $M=3$.}
\end{figure}

In the ETM, as introduced above, we have introduced $M$ {\it
identical} subsystems. If this corresponded to the real world of
BMLJ systems, one would expect a very narrow distribution $p(z)$.
The width of $p(z)$ therefore implies that the assumption of
identical subsystems cannot hold. In any event, it is natural to
assume that at a given time the set of subsystems displays
different sizes;  see Fig.\ref{fig9} for a simple sketch.

To understand the effect of these fluctuations on the waiting time
distribution one may consider the case of $M$ identical subsystems
for fixed total energy $e$. The maximum waiting time of the total
system is realized if by chance all $M$ subsystems have the energy
$e/M$ and the corresponding waiting time $\tau_e$. Otherwise there
is at least one subsystem with higher energy and (on average)
shorter waiting time as compared to $\tau_e$. Thus the presence of
the long-time wing of the waiting time distribution at constant
energy is critically related to the presence of subsystems with
identical waiting times.  Allowing for size fluctuations it is
likely that one subsystem is smaller than the other. Since on
average smaller systems are faster than larger systems (see, e.g.,
Tab.2) one can expect that for a given total energy $e$ this
subsystem and thus the total system has a waiting time much
smaller than $\tau_e$. Therefore size fluctuations strongly reduce
the probability to have a situation with a very long  waiting time
for the whole system. As a consequence, these fluctuations
strongly suppress the long-time tail of the waiting time
distribution, thereby reducing higher moments of this
distribution. One may speculate that this is the reason that the
fluctuations for the BMLJ are somewhat smaller than expected from
the average waiting times. Of course, additionally also the value
of $M$ may somewhat differ for different MB.

\subsection{Implications for dynamic heterogeneities}

The presence of short and long waiting times can be directly
related to the concept of dynamic heterogeneities, observed in
many experiments. As discussed above, a relevant measure for the
presence of dynamic heterogeneities is the value of $S(T)$. On the
level of the elementary subsystem in the ETM the dynamic
heterogeneity is exclusively related to the presence of different
trap depths which, via Eq.\ref{eqtau}, gives rise to a broad
distribution of waiting times. For the BMLJ system with 65 or more
particles, i.e. (in the language of the ETM) for a superposition
of a few elementary subsystems, the dynamic heterogeneities can be
formally divided into two parts. First, already at constant total
energy $e$ one has a range of different waiting times as reflected
by $S_0(T) > 1$ (for $M>1$). Second, the average waiting time
$\langle \tau(e,T)\rangle$ depends on energy.  Comparing the size
of $\sqrt{S_0(T=0.45)} \approx 3$ with the range of the energy
dependence of $\langle \tau(e,T)\rangle$ it becomes obvious that
for 65 particles the dynamic heterogeneities are mainly determined
by the energy dependence of the average waiting time.

\begin{figure}
  \includegraphics[width=8.0cm]{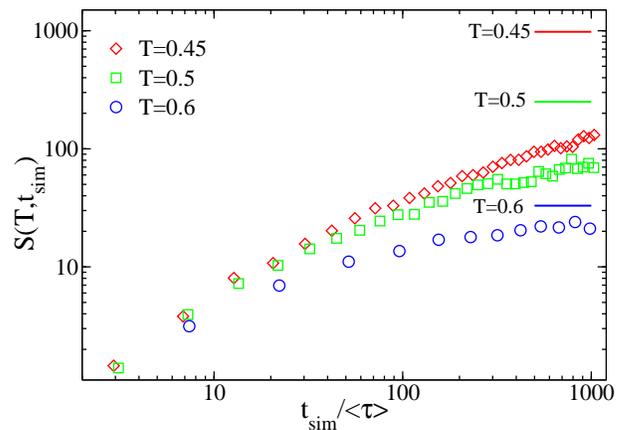}
  \caption{\label{fig10} The dependence of the normalized variance $S(T,t_{sim})$
    on simulation time $t_{sim}$ and temperature for the BMLJ system. The straight lines
    are the estimates of the ETM for the respective limiting values. The estimates for the ETM have
    been scaled by a factor, reflecting the deviations for
    $S_0(T)$
    between the ETM and the BMLJ system.}
\end{figure}

It is possible to characterize the properties of dynamic
heterogeneities somewhat closer. As discussed before, the dynamic
heterogeneities or, equivalently, the width of the waiting time
distribution is captured by the normalized variance $S(T)$ (see
Eq.\ref{defvar}). In the limit of very long simulation times
$t_{sim}$ as compared to the range of waiting times, the
equilibrium value of $S(T)$ can be extracted from the simulations.
How to find an appropriate way of extracting corresponding
information at shorter simulation times?  As already discussed in
the context of the determination of $\langle \tau(T) \rangle$ it
is convenient to analyse the number of MB transitions. Beyond the
determination of the average number of MB transitions one can
estimate its variance $\langle ( n - \langle n \rangle)^2 \rangle
$ by comparing the number of MB transitions during a fixed time
$t_{sim}$ of several independent simulations. Then one may define
\begin{equation}
\label{n2inv} S(T,t_{sim}) = \frac{\langle ( n - \langle n
\rangle)^2 \rangle}{\langle n \rangle^2} \frac{t_{sim}}{\langle
\tau \rangle}.
\end{equation}
As shown in Appendix II in the limit of long $t_{sim}$ the
quantity $S(T,t_{sim})$ approaches the equilibrium value of
$S(T)$. For shorter times $S(T,t_{sim})$ reflects the normalized
variance of that part of the waiting time distribution which can
be accessed on the time scale of the simulation $t_{sim}$. Thus
any dependence of $S(T,t_{sim})$ on $t_{sim}$ indicates that the
finite simulation time introduces an artificial cutoff for the
waiting time distribution. As shown in Fig.\ref{fig10} for $T \le
0.5$ the value of $S(T,t_{sim})$ for the BMLJ system still shows a
strong increase  for $t_{sim} \approx 1000 \langle \tau \rangle
\approx 20 \tau_\alpha$ which already corresponds to relatively
long simulation times.

It may instructive to compare the long-time limit of
$S(T,t_{sim})$ for the BMLJ system with the predictions of the
ETM. We have estimated the latter value by taking (using $M=3$)
$S(T) \cdot [S_{0,BMLJ}(T)/S_{0,ETM}(T)]$. The second factor
corrects for the fact that the actual dynamic heterogeneities of
traps with identical energy are smaller than the predictions of
the ETM; see Fig.\ref{fig6}. The results are included in
Fig.\ref{fig10}.
 For the highest temperature $T=0.6$ the long-time limit of
$S(T,t_{sim})$ can be reached for $t_{sim} \approx 1000 \langle
\tau \rangle$.  Good agreement with the prediction of the ETM is
observed. The large estimates of $S(T)$ within the ETM for the two
lower temperatures reflect the broad dynamic heterogeneity at low
temperatures. They only fully show up in the limit of very long
simulation times.

This result has interesting consequences for the planning of
simulations. The value of $S(T,t_{sim})$ is a measure how
precisely the true value of $\langle \tau \rangle$ can be
extracted from the simulation. More specifically, from $L$
independent runs the value of  $\langle \tau \rangle$ can be
extracted with a precision of $\sqrt{(S(T,t_{sim}))/(L
t_{sim}/\langle \tau \rangle)}$. Here we have assumed that a
simulation of length $t_{sim}$ implies the drawing of
$t_{sim}/\langle \tau \rangle$ independent waiting times. If
$S(T,t_{sim})$ does not depend on $t_{sim}$ one could either
perform a long simulation or several short simulations to obtain
the same quality of $\langle \tau \rangle$. Due to the
time-dependence of $S(T,t_{sim})$  it is more favorable to choose
$t_{sim}$ relatively short and thus perform many independent runs
for fixed total computer time. Alternatively, one can also perform
a few simulations for a correspondingly larger system and obtain
the same quality for the average waiting time. Since the diffusion
constant is directly related to $\langle \tau \rangle$ the same
conclusions hold for the determination of the diffusion constant.

Recently, the variance of the mean square displacement $\langle (
r^2 - \langle r^2 \rangle)^2 \rangle/\langle r^2 \rangle^2$ has
been obtained from comparison of independent simulation runs
\cite{Emilia04}. On a superficial level this ratio resembles the
quantity $\langle ( n - \langle n \rangle)^2 \rangle/\langle n
\rangle^2$ analyzed above. A possible mapping between the
observables $r^2$ and $n$ is further suggested by the validity of
Eq. \ref{eqrw}. Closer analysis, however, reveals that this
mapping is not possible. For a simple random-walker in 1D one has
$p(r^2)dr^2 = \exp(-r^2/\langle r^2 \rangle)dr^2$ \cite{Emilia04}.
In contrast, according to Appendix II the the value of $n$
displays a gaussian distribution around $\langle n \rangle$. As
discussed in \cite{Emilia04} the information content of the
variance of the mean square displacement is more related to the
size of dynamic heterogeneities whereas in the present case we are
sensitive to the distribution of waiting times, i.e. to the degree
of dynamic heterogeneities. In any event, the common idea is to
use independent simulations to extract important information about
the non-trivial dynamic properties of supercooled liquids.

Following the general ideas, underlying the Adam-Gibbs scenario or
alternative descriptions of the physics of supercooled liquids at
low temperatures \cite{Kirkpatrick89,Xia01,Bouchaud04} one might
expect a growing length scale of the cooperativity range. In the
present approach the length scale is reflected by the slope of $T
\ln \langle \tau(e,T) \rangle$. Indeed, some temperature
dependence has been observed, changing the slope by roughly 30 \%
when changing the temperature from $T_c$ (0.45) to $1.33 T_c$
(0.6). In terms of the ETM this implies a weak temperature
dependence of the cooperativity size. Actually, the slopes for $T
=0.45$ and $T=0.5$ were compatible with $M = 3$.

\begin{figure}
  \includegraphics[width=8.0cm]{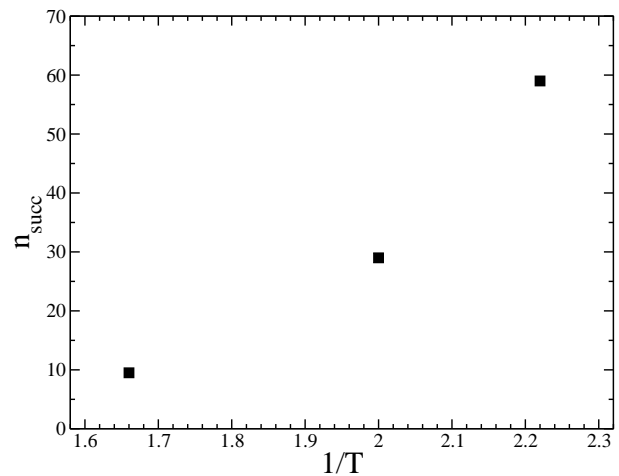}
  \caption{\label{successive} The average number $n_{succ}$ of successive hopping processes
  of the same subensemble in the ETM with $M=3$ for different temperatures.  }
\end{figure}

It may be instructive to compare this weak temperature dependence
with that of other observables which characterize length scales of
dynamic heterogeneities. Analogous behavior is observed for the
average value of the participation number $z$; see Eq.\ref{defz}.
This value is basically constant when comparing $T=0.5$ and
$T=0.435$ \cite{Vogel:2004}. In contrast, a strong
temperature-dependence is observed for quantities which are
determined from the space-time behavior of particles. For example,
the maximum of the generalized susceptibility , characterizing the
size of cooperativity regions, varies by roughly a factor of 5 in
the temperature regime analyzed in this work; see e.g. the
behavior of $\chi_M$ in Ref.\cite{Glotzer00}.  How to rationalize
this apparent discrepancy? In the language of MB, dynamic
heterogeneities have different facets: (i) The spatial range of
particles moving during {\it individual} MB transitions. (ii)
Spatial correlations of {\it subsequent} transitions
\cite{Vogel:2004}. Both aspects are part of the ETM. The first
aspect is directly reflected by the value of $M$. The second
aspect is related to the question how many successive hops are on
average performed by the same subsystem until a different
subsystem relaxes. This value is denoted $n_{succ}$. It can be
easily calculated from simulation of the ETM. As shown in
Fig.\ref{successive} $n_{succ}$ is strongly temperature-dependent
and increases within the temperature regime of interest by a
factor of 6. In the space-time analysis one is sensitive to the
{\it sequence} of fast processes, i.e. the sequence of MB
transitions. Thus one may expect that the increase of the
generalized susceptibility with decreasing temperature is at least
partly due to the increase of $n_{succ}$. On a qualitative level
similar effects have been already reported in \cite{Vogel:2004}
where the formation of macro-strings out of micro-strings has been
observed.

\section{Summary}

Many observations for the BMLJ system are consistent with the
notion that the PEL can be characterized as a superposition of
individual trap-like systems where the individual traps are
identified as MB (and {\it not} as inherent structures). The main
observations are (i) the random-walk nature of the dynamics in
phase space, (ii) the exponential waiting time distribution to
leave MB, (iii) the temperature independent distance of subsequent
MB, (iv) the absence of correlations of subsequent MB waiting
times, and (v) the linear energy-dependence of $T \ln \langle
\tau(e,T) \rangle$ for all temperatures. Analysis of the slightly
temperature-dependent slope of $T \ln \langle \tau(e,T) \rangle$
yields information about the number $M$ of subsystems and the
common barrier height $b$.

From the predictions of this simple model two deviations from the
behavior of the BMLJ system have been observed. First, the energy
correlation function decays faster than expected from the
superposition of traps. This observation directly indicates the
presence of some interaction between these subsystems, expressed
by the dynamic interaction parameter $q$, thereby completing our
definition of the extended trap model (ETM). Second, the dynamic
heterogeneities of MB with constant $e$ were smaller than
expected. This may be a natural consequence of size fluctuations
of the subsystems. In principle it would be possible to include
size fluctuations into the ETM by introducing a new parameter.
This is, however, beyond the scope of the present work.

Our results may be helpful in several ways. The comparison of the
BMLJ dynamics with the predictions of a superposition of
appropriately chosen trap models may elucidate important
properties of the PEL of a prototype glass-forming system.
Furthermore, this analysis may serve as a fresh look onto the
properties of dynamic heterogeneities. Finally, the ETM has the
ability to reproduce also non-equilibrium properties at least on a
semi-quantitative basis. For example it may be feasible to
reproduce  complex aging experiments and to obtain a simple
physical picture of the resulting observations. For this purpose
it is helpful that the relevant length scale for the ETM, as
expressed by the value of $M$, is only weakly temperature
dependent.

{\bf Appendix I}

We consider a simple system which can be in different states with
weights $\varphi_i$. The relaxation process in each state is
characterized by an exponential waiting time distribution with
average $\tau_i$. Thus the resulting average waiting time is given
by $\langle \tau \rangle = \sum_i \varphi_i \tau_i$.  In
equilibrium the probability to be in state $i$ is given by $p_i =
\varphi_i \tau_i / \langle \tau \rangle$.  First, we consider a
simulation of time $t_{sim}$ which is much smaller than any of the
$\tau_i$. In particular one has $t_{sim} \ll \langle \tau
\rangle$. The goal is to estimate the value of $\langle \tau
\rangle$ from the outcome of the simulation. As described in the
main text we may proceed in two different ways. One option is to
determine the average waiting time. Since $t_{sim}$ is very small
at most a single relaxation process will be observed. For
calculating the average waiting time (in the ensemble average) the
most convenient way would be to restrict oneself to those
instances where at least a single relaxation process has occurred.
Due to the smallness of $t_{sim}$ it may occur at any time with
equal probability so that the average waiting time will be
$t_{sim}/2$. This value is, of course, much smaller than the true
average waiting time $\langle \tau \rangle$. Actually, only if
$t_{sim}$ is much larger than all $\tau_i$ a consistent estimate
is possible. The second option is to determine the average number
$\langle n \rangle$ of relaxation processes. Weighting the initial
state with its probability $p_i$, i.e. assuming equilibrium
conditions, and taking into account that the chance of a
relaxation process during time $t_{sim}$ is given by
$t_{sim}/\tau_i$ one ends up with
\begin{equation}
\langle n \rangle = \sum_i p_i \frac{t_{sim}}{\tau_i}= \sum_i
\varphi_i \frac{t_{sim}}{\langle \tau \rangle} =
\frac{t_{sim}}{\langle \tau \rangle}.
\end{equation}
Thus $t_{sim}/\langle n \rangle$ is an exact estimate of the
average waiting time $\langle \tau \rangle$. This result also
holds for larger values of $t_{sim}$ because the large time
interval can be decomposed into small intervals and the average
value of $\langle n \rangle$ will simply scale with the number of
small time intervals such that $\langle n \rangle =
t_{sim}/\langle \tau \rangle$ will hold for arbitrary $t_{sim}$.

{\bf Appendix II}

Of conceptual interest is the variance $\sigma^2_\tau$ of the
waiting time distribution. We show that it is related to the
variance $\sigma^2_n$, obtained from a set of independent runs.
According to the central limit theorem the probability that it
takes a time $t$ to perform $n$ jumps is given by
\begin{equation}
p(t|n) \propto \exp \left(\frac{-(t-n\langle \tau \rangle)^2}{2n
\sigma_\tau^2} \right).
\end{equation}
Here we used that successive jumps are independent from each other
so that we can consider a drawing of n independent elements from
the waiting time distribution. Fixing the value of $t$ as
$t_{sim}$ this probability can be reinterpreted as the probability
$p(n|t)$ that exactly $n$ jumps occur during time $t_{sim}$. Thus
one obtains
\begin{equation}
p(n|t_{sim}) \propto \exp \left ( \frac{-(n - t_{sim}/\langle \tau
\rangle)^2}{2n \sigma_\tau^2/\langle \tau \rangle^2} \right).
\end{equation}
In the limit of large $t_{sim}$ the n-dependence of the
normalization factor can be neglected. Furthermore the factor $n$
in the denominator of the exponential can be substituted by
$\langle n \rangle = t_{sim}/\langle \tau \rangle$.  Therefore the
variance of $n$, i.e. $\sigma_n$, can be written as
\begin{equation}
\frac{\sigma_n^2}{\langle n \rangle^2} = \frac{
\sigma_\tau^2}{\langle \tau \rangle \langle n \rangle} =
\frac{\langle \tau \rangle}{t_{sim}} \frac{ \sigma_\tau^2}{\langle
\tau \rangle^2}
\end{equation}
This directly leads to Eq.\ref{n2inv}.

\begin{acknowledgments}
We like to thank M. Vogel for helpful comments on the manuscript and the
International Graduate School of Chemistry for funding.
\end{acknowledgments}
\bibliography{paperv0.4}

\begin{thebibliography}{56}
\expandafter\ifx\csname natexlab\endcsname\relax\def\natexlab#1{#1}\fi
\expandafter\ifx\csname bibnamefont\endcsname\relax
  \def\bibnamefont#1{#1}\fi
\expandafter\ifx\csname bibfnamefont\endcsname\relax
  \def\bibfnamefont#1{#1}\fi
\expandafter\ifx\csname citenamefont\endcsname\relax
  \def\citenamefont#1{#1}\fi
\expandafter\ifx\csname url\endcsname\relax
  \def\url#1{\texttt{#1}}\fi
\expandafter\ifx\csname urlprefix\endcsname\relax\def\urlprefix{URL }\fi
\providecommand{\bibinfo}[2]{#2}
\providecommand{\eprint}[2][]{\url{#2}}

\bibitem[{\citenamefont{Goldstein}(1969)}]{Goldstein:1969}
\bibinfo{author}{\bibfnamefont{M.}~\bibnamefont{Goldstein}},
  \bibinfo{journal}{J. Chem. Phys.} \textbf{\bibinfo{volume}{51}},
  \bibinfo{pages}{3728} (\bibinfo{year}{1969}).

\bibitem[{\citenamefont{Debenedetti and Stillinger}(2001)}]{Debenedetti:217}
\bibinfo{author}{\bibfnamefont{P.~G.} \bibnamefont{Debenedetti}}
  \bibnamefont{and} \bibinfo{author}{\bibfnamefont{F.~H.}
  \bibnamefont{Stillinger}}, \bibinfo{journal}{Nature}
  \textbf{\bibinfo{volume}{410}}, \bibinfo{pages}{259} (\bibinfo{year}{2001}).

\bibitem[{\citenamefont{Wales}(2003)}]{Wales:2003}
\bibinfo{author}{\bibfnamefont{D.~J.} \bibnamefont{Wales}},
  \emph{\bibinfo{title}{Energy landscapes}} (\bibinfo{publisher}{Cambridge
  University Press}, \bibinfo{year}{2003}).

\bibitem[{\citenamefont{Stillinger and Weber}(1982)}]{Stillinger:1982}
\bibinfo{author}{\bibfnamefont{F.~H.} \bibnamefont{Stillinger}}
  \bibnamefont{and} \bibinfo{author}{\bibfnamefont{T.~A.} \bibnamefont{Weber}},
  \bibinfo{journal}{Phys Rev. A} \textbf{\bibinfo{volume}{25}},
  \bibinfo{pages}{978} (\bibinfo{year}{1982}).

\bibitem[{\citenamefont{Sciortino et~al.}(1999)\citenamefont{Sciortino, Kob,
  and Tartaglia}}]{Sciortino:1999}
\bibinfo{author}{\bibfnamefont{F.}~\bibnamefont{Sciortino}},
  \bibinfo{author}{\bibfnamefont{W.}~\bibnamefont{Kob}}, \bibnamefont{and}
  \bibinfo{author}{\bibfnamefont{P.}~\bibnamefont{Tartaglia}},
  \bibinfo{journal}{Phys Rev. Lett.} \textbf{\bibinfo{volume}{83}},
  \bibinfo{pages}{3214} (\bibinfo{year}{1999}).

\bibitem[{\citenamefont{B{\"u}chner and Heuer}(1999)}]{Buchner:193}
\bibinfo{author}{\bibfnamefont{S.}~\bibnamefont{B{\"u}chner}} \bibnamefont{and}
  \bibinfo{author}{\bibfnamefont{A.}~\bibnamefont{Heuer}},
  \bibinfo{journal}{Phys. Rev. E} \textbf{\bibinfo{volume}{60}},
  \bibinfo{pages}{6507} (\bibinfo{year}{1999}).

\bibitem[{\citenamefont{Mossa et~al.}(2002)\citenamefont{Mossa, {La~Nave},
  Stanley, Donati, Sciortino, and Tartaglia}}]{Mossa:284}
\bibinfo{author}{\bibfnamefont{S.}~\bibnamefont{Mossa}},
  \bibinfo{author}{\bibfnamefont{E.}~\bibnamefont{{La~Nave}}},
  \bibinfo{author}{\bibfnamefont{H.~E.} \bibnamefont{Stanley}},
  \bibinfo{author}{\bibfnamefont{C.}~\bibnamefont{Donati}},
  \bibinfo{author}{\bibfnamefont{F.}~\bibnamefont{Sciortino}},
  \bibnamefont{and}
  \bibinfo{author}{\bibfnamefont{P.}~\bibnamefont{Tartaglia}},
  \bibinfo{journal}{Phys. Rev. E} \textbf{\bibinfo{volume}{65}},
  \bibinfo{pages}{1205} (\bibinfo{year}{2002}).

\bibitem[{\citenamefont{B\"{u}chner and Heuer}(1999)}]{Buechner:1999}
\bibinfo{author}{\bibfnamefont{S.}~\bibnamefont{B\"{u}chner}} \bibnamefont{and}
  \bibinfo{author}{\bibfnamefont{A.}~\bibnamefont{Heuer}},
  \bibinfo{journal}{Phys. Rev. E} \textbf{\bibinfo{volume}{60}},
  \bibinfo{pages}{6507} (\bibinfo{year}{1999}).

\bibitem[{\citenamefont{{La~Nave}
  et~al.}(2002{\natexlab{a}})\citenamefont{{La~Nave}, Stanley, and
  Sciortino}}]{LaNave:265}
\bibinfo{author}{\bibfnamefont{E.}~\bibnamefont{{La~Nave}}},
  \bibinfo{author}{\bibfnamefont{H.~E.} \bibnamefont{Stanley}},
  \bibnamefont{and}
  \bibinfo{author}{\bibfnamefont{F.}~\bibnamefont{Sciortino}},
  \bibinfo{journal}{Phys. Rev. Lett.} \textbf{\bibinfo{volume}{88}},
  \bibinfo{pages}{5501} (\bibinfo{year}{2002}{\natexlab{a}}).

\bibitem[{\citenamefont{Starr et~al.}(2001)\citenamefont{Starr, Sastry,
  {La~Nave}, Scala, Stanley, and Sciortino}}]{Starr:140}
\bibinfo{author}{\bibfnamefont{F.~W.} \bibnamefont{Starr}},
  \bibinfo{author}{\bibfnamefont{S.}~\bibnamefont{Sastry}},
  \bibinfo{author}{\bibfnamefont{E.}~\bibnamefont{{La~Nave}}},
  \bibinfo{author}{\bibfnamefont{A.}~\bibnamefont{Scala}},
  \bibinfo{author}{\bibfnamefont{H.~E.} \bibnamefont{Stanley}},
  \bibnamefont{and}
  \bibinfo{author}{\bibfnamefont{F.}~\bibnamefont{Sciortino}},
  \bibinfo{journal}{Phys. Rev. E} \textbf{\bibinfo{volume}{63}},
  \bibinfo{pages}{1201} (\bibinfo{year}{2001}).

\bibitem[{\citenamefont{Scala et~al.}(2000)\citenamefont{Scala, Starr,
  {La~Nave}, Sciortino, and Stanley}}]{Scala:71}
\bibinfo{author}{\bibfnamefont{A.}~\bibnamefont{Scala}},
  \bibinfo{author}{\bibfnamefont{F.~W.} \bibnamefont{Starr}},
  \bibinfo{author}{\bibfnamefont{E.}~\bibnamefont{{La~Nave}}},
  \bibinfo{author}{\bibfnamefont{F.}~\bibnamefont{Sciortino}},
  \bibnamefont{and} \bibinfo{author}{\bibfnamefont{H.~E.}
  \bibnamefont{Stanley}}, \bibinfo{journal}{Nature}
  \textbf{\bibinfo{volume}{406}}, \bibinfo{pages}{166} (\bibinfo{year}{2000}).

\bibitem[{\citenamefont{Keyes}(1994)}]{Keyes:94}
\bibinfo{author}{\bibfnamefont{T.}~\bibnamefont{Keyes}}, \bibinfo{journal}{J.
  Chem. Phys.} \textbf{\bibinfo{volume}{101}}, \bibinfo{pages}{5081}
  (\bibinfo{year}{1994}).

\bibitem[{\citenamefont{Heuer}(1997)}]{Heuer:1997}
\bibinfo{author}{\bibfnamefont{A.}~\bibnamefont{Heuer}},
  \bibinfo{journal}{Phys. Rev. Lett.} \textbf{\bibinfo{volume}{78}},
  \bibinfo{pages}{4051} (\bibinfo{year}{1997}).

\bibitem[{\citenamefont{Broderix et~al.}(2000)\citenamefont{Broderix,
  Bhattacharya, Cavagna, Zippelius, and Giardina}}]{Broderix:228}
\bibinfo{author}{\bibfnamefont{K.}~\bibnamefont{Broderix}},
  \bibinfo{author}{\bibfnamefont{K.~K.} \bibnamefont{Bhattacharya}},
  \bibinfo{author}{\bibfnamefont{A.}~\bibnamefont{Cavagna}},
  \bibinfo{author}{\bibfnamefont{A.}~\bibnamefont{Zippelius}},
  \bibnamefont{and} \bibinfo{author}{\bibfnamefont{I.}~\bibnamefont{Giardina}},
  \bibinfo{journal}{Phys. Rev. Lett.} \textbf{\bibinfo{volume}{85}},
  \bibinfo{pages}{5360} (\bibinfo{year}{2000}).

\bibitem[{\citenamefont{Schr{\o}der et~al.}(2000)\citenamefont{Schr{\o}der,
  Sastry, Dyre, and Glotzer}}]{Schroder:210}
\bibinfo{author}{\bibfnamefont{T.~B.} \bibnamefont{Schr{\o}der}},
  \bibinfo{author}{\bibfnamefont{S.}~\bibnamefont{Sastry}},
  \bibinfo{author}{\bibfnamefont{J.~C.} \bibnamefont{Dyre}}, \bibnamefont{and}
  \bibinfo{author}{\bibfnamefont{S.~C.} \bibnamefont{Glotzer}},
  \bibinfo{journal}{J. Chem. Phys.} \textbf{\bibinfo{volume}{112}},
  \bibinfo{pages}{9834} (\bibinfo{year}{2000}).

\bibitem[{\citenamefont{{La~Nave} et~al.}(2001)\citenamefont{{La~Nave}, Scala,
  Starr, Stanley, and Sciortino}}]{Scala:376}
\bibinfo{author}{\bibfnamefont{E.}~\bibnamefont{{La~Nave}}},
  \bibinfo{author}{\bibfnamefont{A.}~\bibnamefont{Scala}},
  \bibinfo{author}{\bibfnamefont{F.~W.} \bibnamefont{Starr}},
  \bibinfo{author}{\bibfnamefont{H.~E.} \bibnamefont{Stanley}},
  \bibnamefont{and}
  \bibinfo{author}{\bibfnamefont{F.}~\bibnamefont{Sciortino}},
  \bibinfo{journal}{Phys. Rev. E} \textbf{\bibinfo{volume}{64}},
  \bibinfo{pages}{6102} (\bibinfo{year}{2001}).

\bibitem[{\citenamefont{Angelani et~al.}(2002)\citenamefont{Angelani, Leonardo,
  Ruocco, Scala, and Sciortino}}]{Angelani:315}
\bibinfo{author}{\bibfnamefont{L.}~\bibnamefont{Angelani}},
  \bibinfo{author}{\bibfnamefont{R.~D.} \bibnamefont{Leonardo}},
  \bibinfo{author}{\bibfnamefont{G.}~\bibnamefont{Ruocco}},
  \bibinfo{author}{\bibfnamefont{A.}~\bibnamefont{Scala}}, \bibnamefont{and}
  \bibinfo{author}{\bibfnamefont{F.}~\bibnamefont{Sciortino}},
  \bibinfo{journal}{J. Chem. Phys.} \textbf{\bibinfo{volume}{116}},
  \bibinfo{pages}{10297} (\bibinfo{year}{2002}).

\bibitem[{\citenamefont{Wales and Doye}(2001)}]{Wales:213}
\bibinfo{author}{\bibfnamefont{D.~J.} \bibnamefont{Wales}} \bibnamefont{and}
  \bibinfo{author}{\bibfnamefont{J.~P.~K.} \bibnamefont{Doye}},
  \bibinfo{journal}{Phys. Rev. B} \textbf{\bibinfo{volume}{63}},
  \bibinfo{pages}{214204} (\bibinfo{year}{2001}).

\bibitem[{\citenamefont{Denny et~al.}(2003)\citenamefont{Denny, Reichman, and
  Bouchaud}}]{Denny:2003}
\bibinfo{author}{\bibfnamefont{R.~A.} \bibnamefont{Denny}},
  \bibinfo{author}{\bibfnamefont{D.~R.} \bibnamefont{Reichman}},
  \bibnamefont{and} \bibinfo{author}{\bibfnamefont{J.-P.}
  \bibnamefont{Bouchaud}}, \bibinfo{journal}{Phys. Rev. Lett.}
  \textbf{\bibinfo{volume}{90}}, \bibinfo{pages}{025503}
  (\bibinfo{year}{2003}).

\bibitem[{\citenamefont{Doliwa and Heuer}(2003{\natexlab{a}})}]{Doliwa:2003a}
\bibinfo{author}{\bibfnamefont{B.}~\bibnamefont{Doliwa}} \bibnamefont{and}
  \bibinfo{author}{\bibfnamefont{A.}~\bibnamefont{Heuer}},
  \bibinfo{journal}{Phys. Rev. E} \textbf{\bibinfo{volume}{67}},
  \bibinfo{pages}{031506} (\bibinfo{year}{2003}{\natexlab{a}}).

\bibitem[{\citenamefont{Doliwa and Heuer}(2003{\natexlab{b}})}]{Doliwa:2003b}
\bibinfo{author}{\bibfnamefont{B.}~\bibnamefont{Doliwa}} \bibnamefont{and}
  \bibinfo{author}{\bibfnamefont{A.}~\bibnamefont{Heuer}},
  \bibinfo{journal}{Phys. Rev. E} \textbf{\bibinfo{volume}{67}},
  \bibinfo{pages}{030501} (\bibinfo{year}{2003}{\natexlab{b}}).

\bibitem[{\citenamefont{Doliwa and Heuer}(2003{\natexlab{c}})}]{Doliwa:2003c}
\bibinfo{author}{\bibfnamefont{B.}~\bibnamefont{Doliwa}} \bibnamefont{and}
  \bibinfo{author}{\bibfnamefont{A.}~\bibnamefont{Heuer}},
  \bibinfo{journal}{Phys. Rev. Lett.} \textbf{\bibinfo{volume}{91}},
  \bibinfo{pages}{235501} (\bibinfo{year}{2003}{\natexlab{c}}).

\bibitem[{\citenamefont{Nave and Sciortino}(2004)}]{Emilia04}
\bibinfo{author}{\bibfnamefont{E.~L.} \bibnamefont{Nave}} \bibnamefont{and}
  \bibinfo{author}{\bibfnamefont{F.}~\bibnamefont{Sciortino}},
  \bibinfo{journal}{J. Phys. Chem. B} \textbf{\bibinfo{volume}{108}},
  \bibinfo{pages}{19663} (\bibinfo{year}{2004}).

\bibitem[{\citenamefont{{La~Nave}
  et~al.}(2002{\natexlab{b}})\citenamefont{{La~Nave}, Mossa, and
  Sciortino}}]{Emilia:302}
\bibinfo{author}{\bibfnamefont{E.}~\bibnamefont{{La~Nave}}},
  \bibinfo{author}{\bibfnamefont{S.}~\bibnamefont{Mossa}}, \bibnamefont{and}
  \bibinfo{author}{\bibfnamefont{F.}~\bibnamefont{Sciortino}},
  \bibinfo{journal}{Phys. Rev. Lett.} \textbf{\bibinfo{volume}{88}},
  \bibinfo{pages}{5701} (\bibinfo{year}{2002}{\natexlab{b}}).

\bibitem[{\citenamefont{Adam and Gibbs}(1965)}]{adam:39}
\bibinfo{author}{\bibfnamefont{G.}~\bibnamefont{Adam}} \bibnamefont{and}
  \bibinfo{author}{\bibfnamefont{J.~H.} \bibnamefont{Gibbs}},
  \bibinfo{journal}{J. Chem. Phys.} \textbf{\bibinfo{volume}{43}},
  \bibinfo{pages}{139} (\bibinfo{year}{1965}).

\bibitem[{\citenamefont{Sastry}(2001)}]{Sastry:198}
\bibinfo{author}{\bibfnamefont{S.}~\bibnamefont{Sastry}},
  \bibinfo{journal}{Nature} \textbf{\bibinfo{volume}{409}},
  \bibinfo{pages}{164} (\bibinfo{year}{2001}).

\bibitem[{\citenamefont{Giovambattista
  et~al.}(2003)\citenamefont{Giovambattista, Buldyrev, Starr, and
  Stanley}}]{Nicolas:411}
\bibinfo{author}{\bibfnamefont{N.}~\bibnamefont{Giovambattista}},
  \bibinfo{author}{\bibfnamefont{S.~V.} \bibnamefont{Buldyrev}},
  \bibinfo{author}{\bibfnamefont{F.~W.} \bibnamefont{Starr}}, \bibnamefont{and}
  \bibinfo{author}{\bibfnamefont{H.~E.} \bibnamefont{Stanley}},
  \bibinfo{journal}{Phys. Rev. Lett.} \textbf{\bibinfo{volume}{90}},
  \bibinfo{pages}{085506} (\bibinfo{year}{2003}).

\bibitem[{\citenamefont{Stillinger and Debenedetti}(2002)}]{Stillinger}
\bibinfo{author}{\bibfnamefont{F.~H.} \bibnamefont{Stillinger}}
  \bibnamefont{and} \bibinfo{author}{\bibfnamefont{P.~G.}
  \bibnamefont{Debenedetti}}, \bibinfo{journal}{J. Chem. Phys}
  \textbf{\bibinfo{volume}{116}}, \bibinfo{pages}{3353} (\bibinfo{year}{2002}).

\bibitem[{\citenamefont{Dyre}(1995)}]{Dyre95}
\bibinfo{author}{\bibfnamefont{J.}~\bibnamefont{Dyre}}, \bibinfo{journal}{Phys.
  Rev. B} \textbf{\bibinfo{volume}{51}}, \bibinfo{pages}{12276}
  (\bibinfo{year}{1995}).

\bibitem[{\citenamefont{Monthus and Bouchaud}(1996)}]{Monthus:310}
\bibinfo{author}{\bibfnamefont{C.}~\bibnamefont{Monthus}} \bibnamefont{and}
  \bibinfo{author}{\bibfnamefont{J.~P.} \bibnamefont{Bouchaud}},
  \bibinfo{journal}{J. Phys. A-Math. Gen.} \textbf{\bibinfo{volume}{29}},
  \bibinfo{pages}{3847} (\bibinfo{year}{1996}).

\bibitem[{\citenamefont{Odagaki}(1995)}]{Odagaki95}
\bibinfo{author}{\bibfnamefont{T.}~\bibnamefont{Odagaki}},
  \bibinfo{journal}{Phys. Rev. Lett.} \textbf{\bibinfo{volume}{74}},
  \bibinfo{pages}{2114} (\bibinfo{year}{1995}).

\bibitem[{\citenamefont{Garrahan and Chandler}(2002)}]{Garrahan02}
\bibinfo{author}{\bibfnamefont{J.~P.} \bibnamefont{Garrahan}} \bibnamefont{and}
  \bibinfo{author}{\bibfnamefont{D.}~\bibnamefont{Chandler}},
  \bibinfo{journal}{Phys. Rev. Lett.} \textbf{\bibinfo{volume}{89}},
  \bibinfo{pages}{035704} (\bibinfo{year}{2002}).

\bibitem[{\citenamefont{B{\"u}chner and Heuer}(2000)}]{Buchner:11}
\bibinfo{author}{\bibfnamefont{S.}~\bibnamefont{B{\"u}chner}} \bibnamefont{and}
  \bibinfo{author}{\bibfnamefont{A.}~\bibnamefont{Heuer}},
  \bibinfo{journal}{Phys. Rev. Lett.} \textbf{\bibinfo{volume}{84}},
  \bibinfo{pages}{2168} (\bibinfo{year}{2000}).

\bibitem[{\citenamefont{Stillinger}(1995)}]{Stillinger:333}
\bibinfo{author}{\bibfnamefont{F.~H.} \bibnamefont{Stillinger}},
  \bibinfo{journal}{Science} \textbf{\bibinfo{volume}{267}},
  \bibinfo{pages}{1935} (\bibinfo{year}{1995}).

\bibitem[{\citenamefont{Middleton and Wales}(2001)}]{Middleton:214}
\bibinfo{author}{\bibfnamefont{T.~F.} \bibnamefont{Middleton}}
  \bibnamefont{and} \bibinfo{author}{\bibfnamefont{D.~J.} \bibnamefont{Wales}},
  \bibinfo{journal}{Phys. Rev. B} \textbf{\bibinfo{volume}{64}},
  \bibinfo{pages}{4205} (\bibinfo{year}{2001}).

\bibitem[{\citenamefont{Saksaengwijit et~al.}(2003)\citenamefont{Saksaengwijit,
  Doliwa, and Heuer}}]{Aimorn:2003}
\bibinfo{author}{\bibfnamefont{A.}~\bibnamefont{Saksaengwijit}},
  \bibinfo{author}{\bibfnamefont{B.}~\bibnamefont{Doliwa}}, \bibnamefont{and}
  \bibinfo{author}{\bibfnamefont{A.}~\bibnamefont{Heuer}}, \bibinfo{journal}{J.
  Phys. C: Cond. Mat.} \textbf{\bibinfo{volume}{15}}, \bibinfo{pages}{S1237}
  (\bibinfo{year}{2003}).

\bibitem[{\citenamefont{Reinisch and
  Heuer}(2004{\natexlab{a}})}]{Reinisch:2004a}
\bibinfo{author}{\bibfnamefont{J.}~\bibnamefont{Reinisch}} \bibnamefont{and}
  \bibinfo{author}{\bibfnamefont{A.}~\bibnamefont{Heuer}},
  \bibinfo{journal}{Phys. Rev. B} \textbf{\bibinfo{volume}{70}},
  \bibinfo{pages}{064201} (\bibinfo{year}{2004}{\natexlab{a}}).

\bibitem[{\citenamefont{Reinisch and
  Heuer}(2004{\natexlab{b}})}]{Reinisch:2004b}
\bibinfo{author}{\bibfnamefont{J.}~\bibnamefont{Reinisch}} \bibnamefont{and}
  \bibinfo{author}{\bibfnamefont{A.}~\bibnamefont{Heuer}}, \bibinfo{journal}{J.
  Low Temp. Phys.} \textbf{\bibinfo{volume}{137}}, \bibinfo{pages}{267}
  (\bibinfo{year}{2004}{\natexlab{b}}).

\bibitem[{\citenamefont{Vogel et~al.}(2004)\citenamefont{Vogel, Doliwa, Heuer,
  and Glotzer}}]{Vogel:2004}
\bibinfo{author}{\bibfnamefont{M.}~\bibnamefont{Vogel}},
  \bibinfo{author}{\bibfnamefont{B.}~\bibnamefont{Doliwa}},
  \bibinfo{author}{\bibfnamefont{A.}~\bibnamefont{Heuer}}, \bibnamefont{and}
  \bibinfo{author}{\bibfnamefont{S.~C.} \bibnamefont{Glotzer}},
  \bibinfo{journal}{J. Chem. Phys.} \textbf{\bibinfo{volume}{120}},
  \bibinfo{pages}{4404} (\bibinfo{year}{2004}).

\bibitem[{\citenamefont{Trygubenko and Wales}(2004)}]{Trygubenko:2004}
\bibinfo{author}{\bibfnamefont{S.~A.} \bibnamefont{Trygubenko}}
  \bibnamefont{and} \bibinfo{author}{\bibfnamefont{D.~J.} \bibnamefont{Wales}},
  \bibinfo{journal}{J. Chem. Phys.} \textbf{\bibinfo{volume}{121}},
  \bibinfo{pages}{6689} (\bibinfo{year}{2004}).

\bibitem[{\citenamefont{Osborne and Lacks}(2004)}]{Osborne04}
\bibinfo{author}{\bibfnamefont{M.~J.} \bibnamefont{Osborne}} \bibnamefont{and}
  \bibinfo{author}{\bibfnamefont{D.~J.} \bibnamefont{Lacks}},
  \bibinfo{journal}{J. Phys. Chem. B} \textbf{\bibinfo{volume}{108}},
  \bibinfo{pages}{19619} (\bibinfo{year}{2004}).

\bibitem[{\citenamefont{Keyes and Chowdhary}(2001)}]{Keyes:335}
\bibinfo{author}{\bibfnamefont{T.}~\bibnamefont{Keyes}} \bibnamefont{and}
  \bibinfo{author}{\bibfnamefont{J.}~\bibnamefont{Chowdhary}},
  \bibinfo{journal}{Phys. Rev. E} \textbf{\bibinfo{volume}{64}},
  \bibinfo{pages}{2201} (\bibinfo{year}{2001}).

\bibitem[{\citenamefont{Doliwa and Heuer}(2003{\natexlab{d}})}]{Doliwa:404}
\bibinfo{author}{\bibfnamefont{B.}~\bibnamefont{Doliwa}} \bibnamefont{and}
  \bibinfo{author}{\bibfnamefont{A.}~\bibnamefont{Heuer}}, \bibinfo{journal}{J.
  Phys. C: Cond. Mat.} \textbf{\bibinfo{volume}{15}}, \bibinfo{pages}{S849}
  (\bibinfo{year}{2003}{\natexlab{d}}).

\bibitem[{\citenamefont{Kob and Andersen}(1995)}]{Kob:203}
\bibinfo{author}{\bibfnamefont{W.}~\bibnamefont{Kob}} \bibnamefont{and}
  \bibinfo{author}{\bibfnamefont{H.~C.} \bibnamefont{Andersen}},
  \bibinfo{journal}{Phys. Rev. E} \textbf{\bibinfo{volume}{51}},
  \bibinfo{pages}{4626} (\bibinfo{year}{1995}).

\bibitem[{\citenamefont{G\"{o}tze and Sjogren}(1992)}]{Goetze:1992}
\bibinfo{author}{\bibfnamefont{W.}~\bibnamefont{G\"{o}tze}} \bibnamefont{and}
  \bibinfo{author}{\bibfnamefont{L.}~\bibnamefont{Sjogren}},
  \bibinfo{journal}{Rep. Prog. Phys.} \textbf{\bibinfo{volume}{55}},
  \bibinfo{pages}{241} (\bibinfo{year}{1992}).

\bibitem[{\citenamefont{Chowdhary and Keyes}(2004)}]{Chowdhary04}
\bibinfo{author}{\bibfnamefont{J.}~\bibnamefont{Chowdhary}} \bibnamefont{and}
  \bibinfo{author}{\bibfnamefont{T.}~\bibnamefont{Keyes}}, \bibinfo{journal}{J.
  Phys. Chem. B} \textbf{\bibinfo{volume}{108}}, \bibinfo{pages}{19768}
  (\bibinfo{year}{2004}).

\bibitem[{\citenamefont{Doye et~al.}(1999)\citenamefont{Doye, Miller, and
  Wales}}]{Doye:194}
\bibinfo{author}{\bibfnamefont{J.~P.~K.} \bibnamefont{Doye}},
  \bibinfo{author}{\bibfnamefont{M.~A.} \bibnamefont{Miller}},
  \bibnamefont{and} \bibinfo{author}{\bibfnamefont{D.~J.} \bibnamefont{Wales}},
  \bibinfo{journal}{J. Chem. Phys.} \textbf{\bibinfo{volume}{111}},
  \bibinfo{pages}{8417} (\bibinfo{year}{1999}).

\bibitem[{\citenamefont{Doliwa and Heuer}(2000)}]{Doliwa:spatial}
\bibinfo{author}{\bibfnamefont{B.}~\bibnamefont{Doliwa}} \bibnamefont{and}
  \bibinfo{author}{\bibfnamefont{A.}~\bibnamefont{Heuer}},
  \bibinfo{journal}{Phys. Rev. E} \textbf{\bibinfo{volume}{61}},
  \bibinfo{pages}{6898} (\bibinfo{year}{2000}).

\bibitem[{\citenamefont{Donati et~al.}(1999)\citenamefont{Donati, Glotzer,
  Poole, Kob, and Plimpton}}]{Donati:94}
\bibinfo{author}{\bibfnamefont{C.}~\bibnamefont{Donati}},
  \bibinfo{author}{\bibfnamefont{S.~C.} \bibnamefont{Glotzer}},
  \bibinfo{author}{\bibfnamefont{P.~H.} \bibnamefont{Poole}},
  \bibinfo{author}{\bibfnamefont{W.}~\bibnamefont{Kob}}, \bibnamefont{and}
  \bibinfo{author}{\bibfnamefont{S.~J.} \bibnamefont{Plimpton}},
  \bibinfo{journal}{Phys. Rev. E} \textbf{\bibinfo{volume}{60}},
  \bibinfo{pages}{3107} (\bibinfo{year}{1999}).

\bibitem[{\citenamefont{Perera and Harrowell}(1999)}]{Perera:164}
\bibinfo{author}{\bibfnamefont{D.~N.} \bibnamefont{Perera}} \bibnamefont{and}
  \bibinfo{author}{\bibfnamefont{P.}~\bibnamefont{Harrowell}},
  \bibinfo{journal}{J. Chem. Phys.} \textbf{\bibinfo{volume}{111}},
  \bibinfo{pages}{5441} (\bibinfo{year}{1999}).

\bibitem[{\citenamefont{Tracht et~al.}(1998)\citenamefont{Tracht, Wilhelm,
  Heuer, Feng, Schmidt-Rohr, and Spiess}}]{Tracht}
\bibinfo{author}{\bibfnamefont{U.}~\bibnamefont{Tracht}},
  \bibinfo{author}{\bibfnamefont{M.}~\bibnamefont{Wilhelm}},
  \bibinfo{author}{\bibfnamefont{A.}~\bibnamefont{Heuer}},
  \bibinfo{author}{\bibfnamefont{H.}~\bibnamefont{Feng}},
  \bibinfo{author}{\bibfnamefont{K.}~\bibnamefont{Schmidt-Rohr}},
  \bibnamefont{and} \bibinfo{author}{\bibfnamefont{H.~W.}
  \bibnamefont{Spiess}}, \bibinfo{journal}{Phys. Rev. Lett.}
  \textbf{\bibinfo{volume}{81}}, \bibinfo{pages}{2727} (\bibinfo{year}{1998}).

\bibitem[{\citenamefont{Reinsberg et~al.}(2001)\citenamefont{Reinsberg, Qiu,
  Wilhelm, Spiess, and Ediger}}]{Reinsberg01}
\bibinfo{author}{\bibfnamefont{S.~A.} \bibnamefont{Reinsberg}},
  \bibinfo{author}{\bibfnamefont{X.~H.} \bibnamefont{Qiu}},
  \bibinfo{author}{\bibfnamefont{M.}~\bibnamefont{Wilhelm}},
  \bibinfo{author}{\bibfnamefont{H.~W.} \bibnamefont{Spiess}},
  \bibnamefont{and} \bibinfo{author}{\bibfnamefont{M.~D.}
  \bibnamefont{Ediger}}, \bibinfo{journal}{J. Chem. Phys.}
  \textbf{\bibinfo{volume}{114}}, \bibinfo{pages}{7299} (\bibinfo{year}{2001}).

\bibitem[{\citenamefont{Kirkpatrick et~al.}(1989)\citenamefont{Kirkpatrick,
  Thirumalai, and Wolynes}}]{Kirkpatrick89}
\bibinfo{author}{\bibfnamefont{T.~R.} \bibnamefont{Kirkpatrick}},
  \bibinfo{author}{\bibfnamefont{D.}~\bibnamefont{Thirumalai}},
  \bibnamefont{and} \bibinfo{author}{\bibfnamefont{P.~G.}
  \bibnamefont{Wolynes}}, \bibinfo{journal}{Phys. Rev. A}
  \textbf{\bibinfo{volume}{40}}, \bibinfo{pages}{1045} (\bibinfo{year}{1989}).

\bibitem[{\citenamefont{Xia and Wolynes}(2001)}]{Xia01}
\bibinfo{author}{\bibfnamefont{X.}~\bibnamefont{Xia}} \bibnamefont{and}
  \bibinfo{author}{\bibfnamefont{P.~G.} \bibnamefont{Wolynes}},
  \bibinfo{journal}{Phys. Rev. Lett.} \textbf{\bibinfo{volume}{86}},
  \bibinfo{pages}{5526} (\bibinfo{year}{2001}).

\bibitem[{\citenamefont{Bouchaud and Biroli}(2004)}]{Bouchaud04}
\bibinfo{author}{\bibfnamefont{J.~P.} \bibnamefont{Bouchaud}} \bibnamefont{and}
  \bibinfo{author}{\bibfnamefont{G.}~\bibnamefont{Biroli}},
  \bibinfo{journal}{J. Chem. Phys.} \textbf{\bibinfo{volume}{121}},
  \bibinfo{pages}{7352} (\bibinfo{year}{2004}).

\bibitem[{\citenamefont{Glotzer et~al.}(2000)\citenamefont{Glotzer, Novikov,
  and Schroder}}]{Glotzer00}
\bibinfo{author}{\bibfnamefont{S.}~\bibnamefont{Glotzer}},
  \bibinfo{author}{\bibfnamefont{V.}~\bibnamefont{Novikov}}, \bibnamefont{and}
  \bibinfo{author}{\bibfnamefont{T.}~\bibnamefont{Schroder}},
  \bibinfo{journal}{J. Chem. Phys.} \textbf{\bibinfo{volume}{112}},
  \bibinfo{pages}{509} (\bibinfo{year}{2000}).

\end{thebibliography}
\end{document}